\documentclass[11pt,a4paper]{article}
\usepackage{epsfig,axodraw}
\usepackage{array,cite,amsmath,amssymb}
\usepackage[flushleft]{threeparttable}
\usepackage{subcaption}
\usepackage{kantlipsum}
\usepackage[usenames,dvipsnames]{xcolor}
\usepackage[breakable, theorems, skins]{tcolorbox}
\tcbset{enhanced}
\DeclareRobustCommand{\mybox}[2][gray!20]{%
\begin{tcolorbox}[   
        breakable,
        left=0pt,
        right=0pt,
        top=0pt,
        bottom=0pt,
        colback=#1,
        colframe=#1,
        width=\dimexpr\textwidth\relax, 
        enlarge left by=0mm,
        boxsep=5pt,
        arc=0pt,outer arc=0pt,
        ]
        #2
\end{tcolorbox}
}
\newcommand{\newc}{\newcommand}
\newc{\gev}{\,GeV}
\newcolumntype{M}[1]{>{\centering\arraybackslash}m{#1}}
\newcolumntype{N}{@{}m{0pt}@{}}
\newc{\mev}{\,MeV}
\newc{\ra}{\rightarrow}
\newc{\rpv}{$\mathrm{\not\!R_p}$}
\newc{\rp}{$\mathrm{R_p}$}
\newc{\real}{\mathcal{R}e}
\newc{\alsm}{{\displaystyle \sum_{\alpha=1,2}}}
\newc{\besm}{{\displaystyle \sum_{\beta=1,2}}}
\newc{\al}{\alpha}
\newc{\sgn}{\mr{sgn}\,}
\newc{\be}{\beta}
\newc{\ga}{\gamma}
\newc{\de}{\delta}
\newc{\sla}{\!\!\!\!\!\not\:\:\!}
\newc{\slab}{\!\!\!\!\!\not\,\,\,}
\newc{\slac}{\!\!\!\!\!\!\!\not\,\,\,\,}
\newc{\met}{$\not\!\!E_T$}
\newc{\cw}{\cos\theta_W}
\newc{\sw}{\sin\theta_W}
\newc{\ssw}{\sin^2\theta_W}
\newc{\ccw}{\cos^2\theta_W}
\newc{\cbe}{\cos\beta}
\newc{\sbe}{\sin\beta}
\newc{\ort}{\frac1{\sqrt{2}}}
\newc{\sh}{\hat{s}}
\newc{\uh}{\hat{u}}
\newc{\tha}{\hat{t}}
\newc{\sa}{\sin\al}
\newc{\ca}{\cos\al}
\newc{\mz}{M_{\mr{Z}}}
\newc{\mw}{M_{\mr{W}}}
\newc{\bv}{$\mathrm{\not\!B}$}
\newc{\lv}{$\mathrm{\not\!L}$}
\newc{\beq}{\begin{equation}}
\newc{\eeq}{\end{equation}}
\newc{\ie}{{\it i.e.\/}\ }
\newc{\lam}{\lambda}
\newc{\cht}{\tilde{\chi}}
\newc{\glt}{\tilde{g}}
\newc{\upt}{\tilde{u}}
\newc{\qkt}{\tilde{q}}
\newc{\elt}{\tilde{\ell}}
\newc{\hgt}{\tilde{H}}
\newc{\nut}{\tilde{\nu}}
\newc{\dnt}{\tilde{d}}
\newc{\ftl}{\mr{\tilde{f}}}
\newc{\psb}{\bar{\psi}}
\newc{\rtt}{2^{1/2}}
\newc{\mut}{\tilde{\mu}}
\newc{\mr}{\mathrm}
\newc{\bath}{\bar{\theta}}
\newc{\tht}{\theta}
\newc{\JC}{{\bf J}}
\newc{\lra}{\longrightarrow}
\newc{\eg}{{\it e.g.\  }}
\newc{\barr}{\begin{eqnarray}}
\newc{\earr}{\end{eqnarray}}
\newc{\me}{\mathcal{M}}
\newc{\dbm}{\partial_\mu}
\newc{\dbmu}{\stackrel{\leftrightarrow\  }{\partial^\mu}}
\newc{\sgm}{\sigma_\mu}
\newc{\captionB}[2]{\caption[{#1}]{{\small {#2}}}}
\newc{\ahref}[2]{#2}
\pdfoutput=1 

\usepackage{jheppub} 
\title{\boldmath Implementation of Angularly Ordered Electroweak Parton Shower in Herwig 7}

\author[a]{M.R. Masouminia,}
\author[a,b]{P. Richardson}

\affiliation[a]{Institute for Particle Physics Phenomenology, Durham University, Durham, UK}
\affiliation[b]{Theoretical Physics Department, CERN, Switzerland}

\emailAdd{mohammad.r.masouminia@durham.ac.uk}
\emailAdd{peter.richardson@durham.ac.uk}

\abstract{ \small
We discuss the necessary steps for implementing an angularly ordered (AO) electroweak (EW) parton shower in \textsf{Herwig 7} multi-purpose event generator. This includes calculating the helicity-dependent \textit{quasi}-collinear EW branching functions that correspond to the full range of final-state EW parton shower, in addition to the initial-state EW gauge vector boson radiations. The results are successfully embedded in the AO \textsf{Herwig 7} shower algorithm and have undergone a set of comprehensive and conclusive performance tests. Furthermore, we have used this EW parton shower algorithm, alongside the existing $QCD+QED$ AO shower, to predict the angular distributions of $W^{\pm}$ bosons in LHC events with high transverse momentum jets. These results are compared against the explicitly generated underlying events as well as the existing ATLAS data to show the effectiveness of the newly implemented $QCD+QED+EW$ AO parton shower scheme.}

\begin{document}
\maketitle
\flushbottom

\section{Introduction}

Since the introduction of the main ideas behind process-independent parton showers \cite{Sjostrand:1985xi,Gottschalk:1986bk,Marchesini:1983bm}, they have been amongst the key components of all multi-purpose event generators for particle physics \cite{Bellm:2017bvx,Bellm:2015jjp,Sjostrand:2014zea,Gleisberg:2008ta}. Nowadays, with the extensive development of their algorithms \cite{Collins:1987cp,Knowles:1987cu,Knowles:1988hu,Knowles:1988vs,Richardson:2018pvo,Gustafson:1986db,Ellis:1986bv,Gustafson:1987rq,Marchesini:1991ch,Sjostrand:1993yb,Gleisberg:2003xi,Nagy:2006kb,Nagy:2007ty,Dinsdale:2007mf,Bauer:2008qh,Schumann:2007mg,Platzer:2009jq,Platzer:2011bc}, the use of computer-generated collinear parton showers has become inseparable from the study of particle physics in high-energy collisions and decays. Although the details of these parton shower implementations significantly differ between different showering programs, all existing general-purpose event generators use QED and QCD initial- and/or final-state parton showers, where an evolution scale parameter controls the flow of particles along all branches of the shower. In the \textsf{Herwig 7} default parton shower algorithm \cite{Bahr:2008pv,Bellm:2015jjp,Bellm:2017bvx}, this evolution scale parameter, $\widetilde{q}$, is regulated by the angular ordering (AO) of successive radiations. 

Whilst the current \textit{QCD+QED} schemes for generating collinear parton showers produce satisfactory results for describing the exiting experimental data up to the current LHC energies (e.g. \cite{ATLAS:2019lpk,ATLAS:2019lbg,Aad:2019ojw,ATLAS:2019gkg}), with the upcoming and inevitable push in the probe energies of the existing and future colliders, one expects to observe non-negligible contributions from the pure electroweak (EW) radiations. This is since, at these very high energies,  heavy particles like EW gauge bosons, Higgs bosons and top quarks may appear as constituents of jets and contribute to radiative corrections, corresponding to the fact that these heavy particles will behave as massless partons as $\widetilde{q}$ grows much larger than their masses. In fact, such an expectation is supported by the LHC observations of Higgs boson production via vector-boson fission \cite{CMS:2019kqw,Aaboud:2018gay}, and has been extensively scrutinized in the recent years \cite{Dawson:2014pea,Han:2014nja,Bellm:2016cks,Darvishi:2019uzp}. Furthermore, it has been hinted that excluding EW real emissions from high-energy processes would cause an imbalance since the corresponding virtual corrections are large and have negative signs \cite{Beenakker:2000kb}. This suggests that the Standard Model (SM) can be considered as an unbroken gauge theory at high energies and one has to treat the real emissions of the EW bosons on equal footing as massless gauge bosons.

The above argument clearly justifies making an effort for introducing a process-independent EW parton shower to correspond to EW splittings at high-energy processes. This would ultimately introduce a well-defined EW enhancement to the production rate of a given underlying event and upgrade the conventional parton shower picture to a \textit{QCD+QED+EW} scheme. A number of theoretical studies have already addressed different parts of EW parton shower \cite{Ciafaloni:2000rp,Ciafaloni:2000gm,Ciafaloni:2005fm,Baur:2006sn} while more complete studies of the details of EW splitting functions for both unbroken and broken SM can be found in \cite{Chen:2016wkt}. Furthermore, some attempts have been made to incorporate EW parton shower in multi-purpose event generators \cite{Chiesa:2013yma,Christiansen:2014kba,Krauss:2014yaa,Mangano:2002ea,Kleiss:2020rcg}. Nevertheless, none of the conventional multi-purpose event generators have yet employed a complete and process-independent EW parton shower to realize a \textit{QCD+QED+EW} level enhancement and treat the full scope of high-energy collinear electroweak physics.

In this paper, we aim to discuss the necessary steps for the implementation of an AO initial-state (IS) and final-state (FS) EW parton shower in \textsf{Herwig 7}\footnote{These modifications will be available to the public with the \textsf{Herwig 7.3} release.}.
To this end, we introduce and derive all viable \textit{quasi}-collinear EW splittings of the SM in their spin-unaveraged forms. This is done for both massless and massive cases, including quark splittings, 
\begin{subequations}
\begin{eqnarray}
q\to q'W^{\pm}, \quad 
q\to qZ^{0}, \quad 
q\to qH,
\label{QEW}
\end{eqnarray}
and gauge boson splittings, 
\begin{eqnarray}
& 
W^{\pm}\to W^{\pm}Z^{0}, \quad 
W^{\pm}\to W^{\pm}\gamma, \quad 
Z^{0}\to W^{+}W^{-}, \quad 
\gamma\to W^{+}W^{-}, 
&
\nonumber \\[0.1in]
&
W^{\pm}\to W^{\pm} H, \quad 
Z^{0} \to Z^{0} H .
&
\label{EWEW}
\end{eqnarray}
\end{subequations}
These newly introduced splitting functions, alongside the $H \to q\bar{q}/WW/ZZ$ and $W/Z \to q \bar{q}$ decay modes that already exist in the \textsf{Herwig 7} decay libraries, would create a satisfactory picture for IS and FS EW radiations in the simulated events. In order to obtain the above splitting functions and to make correct approximations in the {\it quasi}-collinear limit, and for numerical efficiency, we present these results in explicit analytic forms. This is followed by extensive performance tests and an assessment of the effectiveness of employing \textit{QCD+QED+EW} scheme in predicting some high-energy milestone processes. 

One should, however, note that the implemented IS EW parton shower will be limited to (\ref{QEW}) splittings, even though the required EW splitting functions for the full spectrum of the IS EW shower would be the same as the FS case, i.e. (\ref{QEW}) and (\ref{EWEW}). This is because implementing an IS shower follows a backward branching evolution \cite{Bellm:2015jjp} where the appropriated Sudakov form factors depend on the parton distribution functions (PDFs) of the relevant particles. Such involvement, however being relatively straightforward in the QCD and QED IS showers, would be problematic for the case of EW IS shower, since it requires incorporating EW PDFs \cite{Kane:1984bb,Dawson:1984gx} and folding QCD and EW effects into a unified set of evolution equations \cite{Ciafaloni:2005fm,Chen:2016wkt}, which is only relevant in the massless EW theory. On the other hand, the required calculations are numerically expensive while being physically insignificant for the case of IS radiations. Furthermore, the available EW PDFs are not reliable nor accurate enough to be introduced in a general-purpose event generator.

The outline of this paper is as follows. In Section \ref{sec:Kin}, we review the branching kinematics and the parametrizations used in \textsf{Herwig 7}. In Section \ref{sec:SpFn}, all the required splitting functions for the implementation of the EW parton shower have been derived. We particularly separate the transverse and longitudinal components of these splittings and present their massless limits and massive correction terms in the simplest spin-unaveraged forms. The required performance tests and physical analysis for this new shower scheme will be presented in Section \ref{sec:res}, followed by our summary and conclusions in Section \ref{sec:conc}. Finally, in Appendix \ref{sec:HI} we describe the required interface commands for using EW shower in \textsf{Herwig 7}.  

\section{Parton Shower Kinematics}
\label{sec:Kin}

In this section, we will introduce the fundamental shower kinematics and dynamics of \textsf{Herwig 7} in the \textit{quasi}-collinear limit~\cite{Catani:2000ef,Bahr:2008pv}. Generally speaking, the branchings kinematics for all cases relevant to the EW parton shower would be the same. We consider the branching of a particle with mass $m_0$ (the parent particle) to two particles with masses $m_1$ and $m_2$ (the children). Then, in the lab frame, the momentum of the branching particle before the emission is
\begin{equation}
p_{\rm lab} = (\sqrt{{\bf p}^2+m_0^2}; {\bf p}\sin\theta\cos\phi,{\bf p}\sin\theta\sin\phi,{\bf p}\cos\theta),
\end{equation}
where ${\bf p}$ is the magnitude of the particle's 3-momentum. This could be either the on-shell momentum from a previous branching in the shower or a parton from the hard matrix element (ME). For simplicity, we will calculate the branchings in a frame in which the particle is moving along the $z$-axis. Hence, by applying a rotation $R$,
\begin{equation}
R = \left(\begin{array}{ccc}
\cos\theta+(1-\cos\theta)\sin^2\phi& -(1-\cos\theta)\sin\phi\cos\phi&-\sin\theta\cos\phi \\
-(1-\cos\theta)\sin\phi\cos\phi & \cos\theta+(1-\cos\theta)\cos^2\phi & -\sin\theta\sin\phi \\
\sin\theta\cos\phi & \sin\theta\sin\phi & \cos\theta \\		
\end{array} \right),
\end{equation}
the momentum of the branching particle becomes
\begin{equation}
  p =   (\sqrt{{\bf p}^2+m_0^2}; 0,0,{\bf p}).
\end{equation}

\textsf{Herwig 7} uses the Sudakov basis to parametrize the momentum the shower particles,
\begin{equation}
q_i = \alpha_ip+\beta_in +q_{\perp i},
\label{eqn:herwigmomentum}
\end{equation}
where the reference vector $n$ is taken to be
\begin{align}
 n &= (1,0,0,-1).
\end{align}
In this parametrization scheme, the momenta of the children particles are
\begin{subequations}
\begin{eqnarray}
  q_1 &=& zp+\beta_1n +q_\perp , \\
  q_2 &=& (1-z)p+\beta_2n -q_\perp ,
\end{eqnarray}
with $z$ being the light-cone momentum fraction of the first parton and
\begin{eqnarray}
  q_\perp &=& (0;p_\perp\cos\phi,p_\perp\sin\phi,0), \\
  \beta_1 &=& \frac1{2zp\cdot n}\left(p_\perp^2+m_1^2-z^2m_0^2\right), \\
  \beta_2 &=& \frac1{2(1-z)p\cdot n}\left(p_\perp^2+m_2^2-(1-z)^2m_0^2\right).
\end{eqnarray}
Also, the momentum of the off-shell parent particle is
\begin{equation}
  q_0 = p +\beta_0n,
\end{equation}
where
\begin{equation}
  \beta_0 = \beta_1+\beta_2 = \frac1{2p\cdot n}\left(\frac{p_\perp^2}{z(1-z)}+\frac{m_1^2}z+\frac{m_2^2}{1-z}-m_0^2\right),
\end{equation}
such that the virtuality of the branching parton is
\begin{equation}
  q_0^2 = \frac{p_\perp^2}{z(1-z)}+\frac{m_1^2}z+\frac{m_2^2}{1-z}.
\end{equation}
\end{subequations}

We need to evaluate the branchings in the {\it quasi}-collinear limit in which we take the masses and
transverse momentum to zero while keeping the ratio of the masses to the transverse momentum fixed.
Practically this is most easily achieved by rescaling the masses and transverse momentum by a parameter
$\lambda$ and expanding in $\lambda$, {\it i.e.}
\begin{equation}
 m_i \to \lambda m_i, \quad p_\perp \to \lambda p_\perp.
\end{equation}
Using the above kinematics, we can calculate the spinors and the polarization vectors of the incoming and outgoing particles, which are in turn used to derive explicit forms of the EW splitting functions that correspond to the splittings (\ref{QEW}) and (\ref{EWEW}).  

\section{Splitting Functions}
\label{sec:SpFn}

Assuming a generic splitting $\widetilde{ij} \to i + j$, one can write the differential cross-section of the production of the children particles $i$ and $j$ in the \textit{quasi}-collinear limit as 
\begin{equation}
d \sigma_{i+j} \simeq 
{\alpha_{\rm int}(\tilde{q}^2) \over 2 \pi} {d\tilde{q}^2 \over \tilde{q}^2} dz \;
P_{\widetilde{ij} \to i + j}(z,\tilde{q}) \; d \sigma_{\widetilde{ij}},
\label{collaprox}
\end{equation}
with $\alpha_{\rm int}$ as the relevant running coupling constant and $P_{\widetilde{ij} \to i + j}(z,\tilde{q})$ being the splitting function of $\widetilde{ij} \to i + j$ branching that dependends on the light-cone momentum fraction $z$ and the evolution scale $\tilde{q}$. In \textsf{Herwig 7}, these parameters are defined as \cite{Bahr:2008pv}
\begin{eqnarray}
z = {\alpha_i \over \alpha_{\widetilde{ij}}} = {n \cdot q_i \over n \cdot q_{\widetilde{ij}}}, 
\qquad
\tilde{q}^2 = \left. {q_{\widetilde{ij}}^2 - m_{\widetilde{ij}}^2 \over z(1-z) } \right|_{q_i^2=m_i^2,\; q_j^2=m_j^2}.
\end{eqnarray}
Henceforth, the problem of calculating the rates of successive branchings in a parton shower is reduced to finding all relevant splitting functions. So, in the following subsections, we will derive the required EW splitting functions for the implementation of EW parton shower in \textsf{Herwig 7}.

\subsection{$q \to q' V$ Splitting Functions}
\label{subsec:qqV}

The majority of QED branchings can be obtained from the equivalent QCD splitting by replacing $\alpha_S\to\alpha_{\rm EM}$ and the color factor with the charge squared of the fermion (or the scalar boson) in the branching. This is however not true in the case of EW branchings. There are a number of issues for either radiation from electroweak bosons or in the case of the radiation of electroweak bosons, which is more complicated due to the mass of the gauge boson and in particular, the presence of the additional longitudinal polarization states. Also, in this case, all engaged particles have non-zero masses.

In a $q \to q' V$ branching with $V=W^{\pm}, \; Z^0$, the transverse polarization vectors (\textit{i.e.} $\lambda_2=\pm 1$) of the vector boson are the same as for the gluonic radiation from a quark splitting, {\it i.e.}
\begin{equation}
  \epsilon^\mu_{\lambda_2=\pm 1}(q_2) = \left[0; 
    -\frac{\lambda_2}{\sqrt{2}}\left(1-\frac {p_\perp^2 \lambda^2{\rm e}^{i\lambda_2\phi} \cos\phi}{2p^2\left(1-z\right)^2}\right),
    -\frac{i}{\sqrt{2}} +\frac {\lambda_2p_\perp^2\lambda^2{\rm e}^{i\phi} \sin\phi}{2\sqrt {2}p^2 \left(1-z\right)^2},
   -\frac { \lambda_2p_\perp\lambda{\rm e}^{i\lambda_2\phi} }{\sqrt {2} \left(1-z\right) p}\right]. 
 \label{eqn:qqgepsT}
\end{equation}
On the other hand, the spinors for the incoming fermion are given by
\begin{equation}
u_{\frac12}(p) = \left(\begin{array}{c}
\frac {m_0}{\sqrt{2p}}\lambda\\ 0 \\ \sqrt {2p}\left(1+\frac{m_0^2\lambda^2}{8p^2}\right) \\ 0
\end{array}\right) \ \ \ \ 
u_{-\frac12}(p) = \left(\begin{array}{c}
  0\\ \sqrt {2p}\left(1+\frac {m_0^2\lambda^2}{8p^2}\right)\\ 0 \\\frac{m_0}{\sqrt{2p}}\lambda
\end{array}\right),
\label{eqn:qqguin}
\end{equation}
and the spinors for the outgoing fermion are
  \begin{eqnarray}
	\bar{u}_{\frac12}(q_1)  &=&
	\left[\sqrt {2zp}\left(1+\frac{m_0^2\lambda^2}{8p^2}\right),
	\frac {{\rm e}^{-i\phi} p_\perp\lambda}{\sqrt {2zp}},
	\frac {m_1}{\sqrt {2zp}}\lambda,
	\frac {{\rm e}^{-i\phi}p_\perp m_1{\lambda}^2}{\left[2zp\right]^{3/2}}
	\right], \nonumber \\
	\bar{u}_{-\frac12}(q_1)  &=& \left[
	-\frac { {\rm e}^{i\phi} p_\perp m_1\lambda^2}{\left[2zp\right]^{3/2}},
	\frac{m_1}{\sqrt {2zp}}\lambda,
	-\frac { {\rm e}^{i\phi} p_\perp\lambda}{\sqrt{2zp}},
	\sqrt {2zp}\left(1+\frac {m_0^2\lambda^2}{8zp}\right)
	\right].
	\label{eqn:qqguout}
\end{eqnarray}
In this case, we write the vertex for the interaction of the fermions with the gauge boson as
\begin{equation}
  -i g \left(g_LP_L+g_RP_R\right)\gamma^\mu,
\end{equation}
with an arbitrary overall coupling and separate couplings to the left- and right-handed helicities.
In this notation, the vertex for the interactions of the quarks and gluons would have $g=g_s$ and $g_L=g_R=1$.
The helicity amplitudes for the splitting can then be written as 
\begin{equation}
	H_{q\to q' V}(z,\tilde{q};\lam_0,\lam_1,\lam_2) = g\sqrt{\frac2{\tilde{q}^2_0-m_0^2}} F_{\lam_0,\lam_1,\lam_2}^{q\to q' V},
\end{equation}
where the vertex function takes on the form
\begin{equation}
  F_{\lam_0,\lam_1,\lam_2}^{q\to q' V} = \sqrt{\frac1{2(\tilde{q}^2_0-m_0^2)}}\bar{u}_{\lam_1}(q_1)\left(g_LP_L+g_RP_R\right)\epsilon_{\lam_2}\! \! \! \! \slac \ 
	u_{\lam_0}(q_0).
\end{equation}
These functions are given explicitly in Table~\ref{tab:qqWT}.
\begin{table}
  \begin{center}
  \setlength\extrarowheight{5pt}
    \begin{tabular}{|c|c|c|c|c|c|}
      \hline
       & & \multicolumn{4}{c|}{ $F_{\lam_0,\lam_1,\lam_2}^{q\to q' V}$} \\[5pt]
      \cline{3-6}
      $\lam_0$ & $\lam_1$ & $\lam_2=+$ &$\lam_2=-$  & $\lam_2=0$ &$\lam_2=0^*$  \\[5pt]
      \hline \hline
      + & + & $\frac {g_Rp_\perp}{(1-z)\sqrt{z(\tilde{q}^2-m_0^2)}}$ &
      $-\frac {g_Rp_\perp \sqrt{z}}{(1-z)\sqrt{(\tilde{q}^2-m_0^2)}}$ &
      $\frac {g_Lm_0m_1\left(1-z\right)^2+g_R(p_\perp^2-m_2^2z)}{(1-z)\sqrt {2z(\tilde{q}^2-m_0^2)}m_2}$ &
      $-\frac {g_Rm_2}{1-z}\sqrt{\frac{2z}{(\tilde{q}^2-m_0^2)}}$ \\[5pt]
      \hline
      + & - & $-\frac { \left( g_Lm_0z-g_Rm_1 \right)}{\sqrt{z(\tilde{q}^2-m_0^2)}}$ &
      $0$ &
      $-\frac {\left( g_Lm_0-g_Rm_1 \right) p_\perp}{\sqrt {2z(\tilde{q}^2-m_0^2)}m_2}$ &
      $0$ \\[5pt]
      \hline
      - & + & $0$ &
      $\frac { \left( -g_Rm_0z+g_Lm_1 \right) }{\sqrt{z(\tilde{q}^2-m_0^2)}}$ &
      $-\frac {\left( g_Lm_1-g_Rm_0 \right) p_\perp}{\sqrt {2z(\tilde{q}^2-m_0^2)}m_2}$ &
      $0$ \\[5pt]
      \hline
      - & - & $\frac {g_Lp_\perp \sqrt{z}}{(1-z)\sqrt{(\tilde{q}^2-m_0^2)}}$ &
      $-\frac {g_Lp_\perp}{(1-z)\sqrt{z(\tilde{q}^2-m_0^2)}}$ &
      $\frac{\left(g_Rm_0m_1(1-z)^2+g_L(p_\perp^2-m_2^2z)\right)}{(1-z)\sqrt {2z(\tilde{q}^2-m_0^2)}m_2}$ &
      $-\frac {g_Lm_2}{1-z}\sqrt{\frac{2z}{(\tilde{q}^2-m_0^2)}}$ \\[5pt]
      \hline
    \end{tabular}
  \end{center}
\caption{Spin-unaveraged splitting functions for $q \to q' V$. In addition to the factors given above, each term has a phase $e^{i(\lam_0-\lam_1-\lam_2)\phi'}$, where $\lam_{2}=\pm1$ and $\lam_{0,1}=\pm\frac12$ represent the helicity states of the outgoing EW guage vector boson and the quarks respectivey. The longitudinal polarization states, which are marked as $\lam_2=0^*$, give the relevant components in the Dawson's approach \cite{Dawson:1984gx} and can be used to eliminate singularities that appear in the $p_{\perp} \gg m_2$ limit.}
\label{tab:qqWT}
\end{table}

Now, we can sum over the transverse parts of the $q \to q' V$ helicity amplitudes and write the spin-averaged transverse splitting function as
\begin{eqnarray}
	P^{\rm T}_{q\to q' V}(z,\tilde{q}) &=& \sum_{\lambda_0, \lambda_1 = \pm {1 \over 2}; \lambda_2 = \pm 1} 
	\left| H_{q\to q' V}(z,\tilde{q};\lam_0,\lam_1,\lam_2) \right|^2
	\nonumber \\
	&=&
	\frac {1}{1-z} \Bigg(\frac{\left( {g_L}^2+{g_R}^2\right)}
	2\left[1+z^2+\frac {\left(1-z^2\right)\left({m_0}^2-{m_1}^2\right)-\left(1+ z^2\right)  
	{m_2}^2}{z \left( 1-z \right)^2 \tilde{q}^2}\right] \nonumber\\ 
	&& -2g_Lg_R\frac {m_0m_1}{z {\tilde{q}}^2} \Bigg),
\end{eqnarray}
which reduces to the $q\to qg$ splitting function for $g_L=g_R=1$, $m_0=m_1=m$ and $m_2=0$. Obviously, the spin-averaged transverse splitting functions cannot be used alone to generate the EW shower since the branching probability will depend on the helicity of the particle. For example, only the left-handed helicity will couple to the $W^\pm$ bosons,
\begin{eqnarray}
	P^{\rm T}_{q\to q' V}(z,\tilde{q}) &=& 
	\frac {1}{1-z} \left[  \left( {g_L}^2\rho_{{-1,-1}}+{g_R}^2\rho_{{1,1}} \right)  \left\{  (1+z^2) 
	\left( 1+{\frac {m_0^2(1-z)-m_2^2}{z \left(1-z\right)^2\tilde{q}^2}} \right) 
	\right. \right. \nonumber\\ &-& \left. \left.  
	{\frac {{m_1}^2 \left( z+1 \right) }{z \left( 1-z \right) {\tilde{q}}^2}} \right\} 
	+\frac{m_0^2}{\tilde{q}^2}\left(g_L^2\rho_{{1,1}}+g_R^2\rho_{{-1,-1}}\right)
	-\frac {2m_0m_1g_Lg_R}{z\tilde{q}^2} \left(\rho_{{1,1}}+\rho_{{-1,-1}} \right) \right],
	\nonumber \\
\end{eqnarray}
where $\rho$ is the spin density matrix of the $W^\pm$ boson.

On the other hand, the longitudinal polarization vector of a massive EW gauge boson (\textit{i.e.} $\lambda_2=0$) is
\begin{eqnarray}
 \epsilon^\mu_0(q_2) &=& \left[ \frac {p \left(1-z\right)}{\lambda m_2}
 +\frac{p_\perp^2+m_0^2(1-z)^2-m_2^2}{4p \left(1-z\right) m_2}\lambda;
 \cos\phi\left( \frac {p_\perp}{m_2}-\frac {m_2p_\perp\lambda^2}{2p^2\left(1-z\right)^2}\right),
 \right. \nonumber \\
 && \left. -\sin\phi\left( \frac {p_\perp}{m_2}
 -\frac{m_2p_\perp\lambda^2}{2p^2\left(1-z\right)^2}\right),
 \frac {p\left(1-z\right)}{\lambda m_2}
 -{\frac {p_\perp^2-m_0^2(1-z)^2-m_2^2}{4p \left(1-z\right)m_2}}\lambda,
 \right].
 \nonumber \\
 \label{eqn:qqgepsL}
\end{eqnarray}
If we compute the splitting function using this polarization vector, taking $m_0=m_1=m$ for simplicity, we obtain
\begin{eqnarray}
	P^{\rm L}_{q\to q' V}(z,\tilde{q}) &=& \sum_{\lambda_0, \lambda_1= \pm {1 \over 2}; \lambda_2 =0} 
	\left| H_{q\to q' V}(z,\tilde{q};\lam_0,\lam_1,\lam_2) \right|^2
	\nonumber \\
	&=&
	{1\over 2} \left(g_L^2\rho_{{-1,-1}}+g_R^2\rho_{{1,1}} \right)
	{\frac {\left( {\tilde{q}}^{2}z \left( 1-z \right) ^{2}-2{m_2}^{2} \right) ^{2}}{{m_2}^{2} 
	\left( 1-z \right) ^{3}{\tilde{q}}^{2}}}.
\end{eqnarray}
This form, while perfectly valid in the quasi-collinear limit, presents a problem as it has terms
which grow for $p_\perp\gg m_2$, {\it i.e.} 
\begin{equation}
 P^{\rm L}_{q\to q' V}(z,\tilde{q})
 \stackrel{p_\perp\gg m_2}{\longrightarrow} {1\over 2} \left( {g_L}^{2}\rho_{{-1,-1}}+{g_R}^{2}\rho_{{1,1}} \right){\frac {  {\tilde{q}}^{2}{z}^{2} \left( 1-z \right) }{{m_2}^{2}}}.
\end{equation}
We therefore adopt the Dawson's approach \cite{Dawson:1984gx}, where the piece of the longitudinal polarization vector proportional to its momentum is subtracted, giving
\begin{eqnarray}
 \epsilon^\mu_{0^*}(q_2) &=& \frac{\lambda m_2}{2p\left(1-z\right)}\left[
-1;  
\frac {\lambda\cos\phi p_\perp}{p\left(1-z\right)},
\frac {\lambda\sin\phi p_\perp}{p\left(1-z\right)},
 1\right],
 \label{eqn:qqgepsLD}
\end{eqnarray}
which vanishes for $m_2\to0.$ Using this approach, we can derive the longitudinal polarization as
\begin{eqnarray}
	P^{\rm L}_{q\to q' V}(z,\tilde{q}) &=& \sum_{\lambda_0, \lambda_1= \pm {1 \over 2}; \lambda_2 =0^{*}} 
	\left| H_{q\to q' V}(z,\tilde{q};\lam_0,\lam_1,\lam_2) \right|^2
	\nonumber \\
	&=&
	\left(g_L^2\rho_{{-1,-1}}+g_R^2\rho_{{1,1}}\right) 
	\frac {2m_2^2}{\tilde{q}^2 \left(1-z\right)^3}.
\end{eqnarray}
Now, putting all the pieces together, the splitting function of the $q\to q' V$ branching takes on the following form
\begin{eqnarray}
	P_{q\to q' V}(z,\tilde{q}) &=& P^{\rm T}_{q\to q' V}(z,\tilde{q}) + P^{\rm L}_{q\to q' V}(z,\tilde{q})
	\nonumber \\
	&=& \frac {1}{1-z} \left[  \left( {g_L}^2\rho_{{-1,-1}}+{g_R}^2\rho_{{1,1}} \right)
	\left\{  \left(1+z^2 \right)  \left(1+\frac {m_0^2}{\tilde{q}^2z\left(1-z\right)}\right)
	-\frac{m_1^2\left(1+z\right)}{z\tilde{q}^2\left(1-z\right)}
	\right.  \right. \nonumber\\ &-& \left. \left.
	\frac {m_2^2}{z\tilde{q}^2} \right\}  
	+ \frac {{m_0}^2}{\tilde{q}^2} \left( {g_L}^2\rho_{{1,1}}+{g_R}^2\rho_{{-1,-1}}\right)
	-\frac {2m_0m_1}{z\tilde{q}^2} g_Lg_R\left(\rho_{{1,1}}+\rho_{{-1,-1}} \right) \right],
	\nonumber \\
	\label{P(q->qv)}
\end{eqnarray}
that can be decomposed into massless and massive expressions as
\begin{subequations}
\begin{eqnarray}
	P_{q\to q' V}^{\rm massless}(z,\tilde{q}) 
		&=& \left( {g_L}^2\rho_{{-1,-1}}+{g_R}^2\rho_{{1,1}} \right) \frac {1+z^2}{1-z},
	\label{P(q->qv)ML}
	\\
	P_{q\to q' V}^{\rm massive}(z,\tilde{q}) 
		&=& \frac {1}{1-z} \left[  \left( {g_L}^2\rho_{{-1,-1}}+{g_R}^2\rho_{{1,1}} \right)
	\left\{ \frac {m_0^2 (1+z^2)}{\tilde{q}^2z\left(1-z\right)}
	-\frac{m_1^2\left(1+z\right)}{z\tilde{q}^2\left(1-z\right)}
	- \frac {m_2^2}{z\tilde{q}^2} \right\} 
	\right. \nonumber\\ &+& \left.
	\frac {{m_0}^2}{\tilde{q}^2} \left( {g_L}^2\rho_{{1,1}}+{g_R}^2\rho_{{-1,-1}}\right)
	-\frac {2m_0m_1g_Lg_R}{z\tilde{q}^2} \left(\rho_{{1,1}}+\rho_{{-1,-1}} \right) \right].
\label{P(q->qv)M}
\end{eqnarray}
\end{subequations}
From the Eq. (\ref{P(q->qv)ML}), it can be readily seen that in the massless limit, $P_{q\to q' V}$ reduces to its QCD counterpart for $g_L, \; g_R \to 1$.

\subsection{$q \to q H$ Splitting Function}
\label{subsec:qqH}

The case of Higgs boson radiation from a parent quark is arguably the simplest case in the study of EW branchings. Here, the spinors of the incoming and the outgoing quarks are the same as in the $q \to q' V$ case, \textit{i.e.} the Eqs. (\ref{eqn:qqguin}) and (\ref{eqn:qqguout}) respectively. Furthermore, the vertex coupling for a $q \to q H$ splitting is
\begin{equation}
	-i g {m_0 \over m_W},
\end{equation}
with $g = e/(2 sin \theta_W)$ and $m_W$ being the mass of $W$ gauge boson. This suggests that $q \to q H$ splitting would be suppressed by a factor of $(m_0/m_W)^2$ for the light-quarks where $m_0 \ll m_W$. Therefore, it would be safe to consider only heavy quarks as parent particles of $q \to q H$ splittings.

Once more, the corresponding helicity amplitudes can be written as 
\begin{equation}
	H_{q\to qH}(z,\tilde{q};\lam_0,\lam_1) = g {m_0 \over m_W} \sqrt{\frac2{\tilde{q}^2_0-m_0^2}} F_{\lam_0,\lam_1}^{q\to qH},
\end{equation}
with
\begin{equation}
	F_{\lam_0,\lam_1}^{q\to qH} = \sqrt{\frac1{2(\tilde{q}^2_0-m_0^2)}} \bar{u}_{\lam_1}(q_1) u_{\lam_0}(q_0).
\end{equation}
The explicit forms of $F_{\lam_0,\lam_1}^{q\to qH}$ functions are given in Table~\ref{tab:qqH}.
\begin{table}
  \begin{center}
  \setlength\extrarowheight{5pt}
    \begin{tabular}{|c|c|c|}
    \hline 
      $\lam_0$ & $\lam_1$ & $F_{\lam_0,\lam_1}^{q\to qH}$ \\[5pt]
      \hline \hline
      + & + &  $\frac{m_0 (1+z)}{\sqrt{2 z \left(\tilde{q}^2-m_0^2\right)}}$ \\[5pt] 
 \hline
      + & - &  $-\frac{p_{\perp}}{\sqrt{2 z \left(\tilde{q}^2-m_0^2\right)}}$ \\[5pt]
 \hline
      - & + &  $\frac{p_{\perp}}{\sqrt{2 z \left(\tilde{q}^2-m_0^2\right)}}$ \\[5pt]
 \hline
      - & - &  $\frac{m_0 (1+z)}{\sqrt{2 z \left(\tilde{q}^2-m_0^2\right)}}$ \\[5pt]
      \hline
    \end{tabular}
  \end{center}
\caption{Spin-unaveraged splitting functions for $q \to q H$. In addition to the factors given above, each term has a phase $e^{i(\lam_0-\lam_1)\phi'}$, where $\lam_{0,1}=\pm\frac12$ represent the helicity states of the quarks.}
\label{tab:qqH}
\end{table}

Putting the above parts together, we can write the splitting function for the $q\ra qH$ splitting as
\begin{eqnarray}
	P_{q\to qH}(z,\tilde{q}) &=& \sum_{\lambda_0, \lambda_1= \pm {1 \over 2}} 
	\left| H_{q\to qH}(z,\tilde{q};\lam_0,\lam_1) \right|^2
	\nonumber \\
	&=& g^2 ({m_0 \over m_W})^2 \left[ (1-z)+
	\frac{4 m_0^2 - m_2^2 }{\tilde{q}^2 (1-z) z}
	\right],
\end{eqnarray}
where we take $m_0$ and $m_2$ to be the running masses of the parent heavy quark and the child Higgs boson, respectively.

\subsection{$V\to V' V''$ Splitting Functions}
\label{subsec:VVV}

Each of the relevant particles in this case, the parent or any of the children, could be either a massive or a massless gauge vector boson and may have the corresponding transverse and/or longitudinal polarization vectors. For the parent gauge boson we can write  
\begin{eqnarray}
	&&\epsilon^\mu_{\lambda_0=\pm 1}(p) = \left[ 0,-\frac{\lambda_0}{\sqrt{2}},-\frac{i}{\sqrt{2}},0 \right], 
	\label{polT-0}
	\\
	&&\epsilon^\mu_{0}(p) = \left[ 
	\frac{p}{\lambda  m_0},0,0,\frac{\sqrt{\lambda ^2 m_0^2+p^2}}{\lambda  m_0}	
	\right]. 
	\label{polL-0}
\end{eqnarray}
Again, to avoid the $m_0\to0$ singularities that would emerge from the longitudinal polarization vector, we employ the Dawson's approach and rewrite this vector as 
\begin{eqnarray}
	\epsilon^\mu_{0^*}(p) = \left[ 
	-\frac{\lambda  m_0}{p+\sqrt{\lambda ^2 m_0^2+p^2}},0,0,
	\frac{\lambda  m_0}{p+\sqrt{\lambda ^2 m_0^2+p^2}}
	\right]. 
	\label{polLD-0}
\end{eqnarray}
Furthermore, we can use the polarization vectors (\ref{eqn:qqgepsT}) and (\ref{eqn:qqgepsLD}) for the second child of a $V\to V' V''$ splitting, while applying the transformations $z \to (1-z)$ will reproduce the polarization vectors of the first child. 

With the above information, we can derive the helicity amplitudes for a $V\to V' V''$ splitting as 
\begin{equation}
	H_{V\to V' V''}(z,\tilde{q};\lam_0,\lam_1,\lam_2) = i g\sqrt{\frac2{\tilde{q}^2_0-m_0^2}} F_{\lam_0,\lam_1,\lam_2}^{V\to V' V''},
\end{equation}
where $g=e \tan \theta_W$ when $V',V'' = W^{\pm},Z^0$ and $g=e \tan$ when either $V$ or $V''$ is a photon. For these splittings, the vertex functions take on the form
\begin{equation}
	F_{\lam_0,\lam_1,\lam_2}^{V\to V' V''} = \sqrt{\frac1{2(\tilde{q}^2_0-m_0^2)}}
	\left[ 
	(q_1 \cdot \epsilon_{\lambda_2}^*) (\epsilon_{\lambda_0} \cdot \epsilon_{\lambda_1}^*)
	+ (q_2 \cdot \epsilon_{\lambda_0}^*) (\epsilon_{\lambda_1} \cdot \epsilon_{\lambda_2}^*)
	- (q_2 \cdot \epsilon_{\lambda_1}^*) (\epsilon_{\lambda_0} \cdot \epsilon_{\lambda_2}^*)
	\right],
\end{equation}
with their explicit forms given in Table~\ref{tab:VVV}. Note that in this table, we have left out the $\lambda_i = 0$ terms since most of them produce large and complicated vertex functions that depend on powers of $1/m_i$. In these cases, we have replaced the longitudinal polarization vectors with their counterparts in Dawson's approach and carried on the calculation. 
\begin{table}
  \begin{center}
  \setlength\extrarowheight{5pt}
    \begin{tabular}{|c|c|c|c|c|}
		\hline
		& & \multicolumn{3}{c|}{$F_{\lam_0,\lam_1,\lam_2}^{V\to V' V''}$} \\[5pt]
		\cline{3-5}
		$\lam_0$ & $\lam_1$ & $\lam_2=+$ & $\lam_2=-$ & $\lam_2=0^*$ \\[5pt]
		\hline \hline
		+ & + 
		& $-\frac{p_{\perp}}{\sqrt{\tilde{q}^2(1-z)^3z^3}}$ 
		& $\frac{p_{\perp} \sqrt{(1-z) z}}{\tilde{q} (1-z)^2}$ 
		& $\frac{\sqrt{2} m_2 \sqrt{(1-z) z}}{\tilde{q}(1-z)^2}$ \\[5pt]    
		\hline
		+ & - 
		& $\frac{p_{\perp} \sqrt{1-z}}{\tilde{q}z^3}$ 
		& 0 
		& 0 \\[5pt]    
		\hline
		- & + 
		& 0 
		& $-\frac{p_{\perp} \sqrt{1-z}}{\tilde{q}z^3}$ 
		& 0 \\[5pt]   
		\hline
		- & - 
		& $\frac{p_{\perp} z}{\sqrt{\tilde{q}^2(1-z)^3 z}}$ 
		& $\frac{p_{\perp}}{\tilde{q} \sqrt{(1-z)^3 z^3}}$ 
		& $\frac{\sqrt{2} m_2 \sqrt{(1-z) z}}{\tilde{q}(1-z)^2}$ \\ 
		\hline
		+ & $0^*$ 
		& $-\frac{\sqrt{2} m_1 \sqrt{1- z}}{\tilde{q}\sqrt{z^3}}$ 
		& 0 
		& 0 \\[5pt]
		\hline
		$0^*$ & + 
		& 0 
		& $\frac{\sqrt{2} m_0 (z-1)}{\tilde{q} \sqrt{(1-z)z}}$ 
		& 0 \\[5pt]
		\hline
		- & $0^*$ 
		& 0 
		& $-\frac{\sqrt{2} m_1 \sqrt{1- z}}{\tilde{q}\sqrt{z^3}}$ 
		& 0 \\[5pt]
		\hline
		$0^*$ & - 
		& $\frac{\sqrt{2} m_0 (z-1)}{\tilde{q} \sqrt{(1-z)z}}$ 
		& 0 
		& 0 \\[5pt]
		\hline 
		$0^*$ & $0^*$ 
		& 0 
		& 0 
		& 0 \\[5pt]
		\hline
	\end{tabular}
  \end{center}
\caption{Spin-unaveraged splitting functions for $V \to V' V''$. In addition to the factors given above, each term has a phase $e^{i(\lam_0-\lam_1-\lam_2)\phi'}$, where $\lam_{0,1,2}=\pm,0^*$ represent the helicity states of the incoming. The longitudinal polarization states which are marked as $\lam_2=0^*$ give the relevant components in the Dawson's approach \cite{Dawson:1984gx} and can be used to eliminate singularities that appear in the $p_{\perp} \gg m_i, \; i=0,1,2$ limit.}
\label{tab:VVV}
\end{table}

Having the explicit forms of the $F_{\lam_0,\lam_1,\lam_2}^{V\to V' V''}$ functions, we can identify different helicity configurations of the splitting function as follows
\begin{eqnarray}
	P_{V \to V'V''}^{\rm TTT}(z,\tilde{q}) &=& 
	2(\rho_{-1,-1}+\rho_{1,1}) \left( \frac{1-z(1-z)}{z(1-z)} \right)^2
	\left[ 
	m_{0,t}^2 (1-z) z-m_{1,t}^2 (1-z) \right.
	\nonumber \\	
	&-& \left. m_{2,t}^2 z+(1-z) z
	\right],
\end{eqnarray}
\begin{eqnarray}
	P_{V \to V'V''}^{\rm TTL}(z,\tilde{q}) &=& 
	2(\rho_{-1,-1}+\rho_{1,1}) \left( \frac{z }{1-z} \right)^2 m_{2,t}^2,
\end{eqnarray}
\begin{eqnarray}	
	P_{V \to V'V''}^{\rm TLT}(z,\tilde{q}) &=& 2(\rho_{-1,-1}+\rho_{1,1}) \left( \frac{1-z}{z} \right)^2 m_{1,t}^2,
\end{eqnarray}
\begin{eqnarray}
	P_{V \to V'V''}^{\rm LTT}(z,\tilde{q}) &=& 4 \rho_{0,0} (1-z)^2 m_{0,t}^2,
\end{eqnarray}
\begin{eqnarray}	
	P_{V \to V'V''}^{\rm LLL}(z,\tilde{q}) &=& 2 \rho_{0,0} \left({z \over 1-z}\right)^2 
	{m_{0,t}^2 m_{2,t}^2 \over m_{1,t}^2},
	\label{vvv:LLL}	
\end{eqnarray}
with $m_{i,t}^2 = m_i^2/(\tilde{q}^2z(1-z))$. The TLL, LTL, LLT and LLL parts vanish at Dawson's approach. Having derived all helicity parts, we can simply extract the final forms of the corresponding $V \to V'V''$ splitting functions as:
\begin{eqnarray}
	P_{V \to V'V''}(z,\tilde{q}) &=& P^{\rm TTT} + P^{\rm TTL} + P^{\rm TLT} 
	+ P^{\rm TLL} + P^{\rm LTT} + P^{\rm LTL} + P^{\rm LLT} + P^{\rm LLL},
	\nonumber \\	 
\end{eqnarray}
which can be separated in massless and massive terms as
\begin{eqnarray}
	P_{V \to V'V''}^{\rm massless}(z,\tilde{q}) &=& 2(\rho_{-1,-1}+\rho_{1,1})
	\frac{(1-(1-z) z)^2}{(1-z) z},
	\label{vvv-massless}
	\\	 
	P_{V \to V'V''}^{\rm massive}(z,\tilde{q}) &=& 
	\frac{1}{(1-z) z} (\rho_{-1,-1}+\rho_{1,1}) 
	\left[
	2 m_{0,t}^2 (1-(1-z) z)^2
	- 2 m_{1,t}^2 \left(1-(1-z) z^2\right)
	\right. \nonumber \\
	&-& \left. 2 m_{2,t}^2 \left(1-(1-z)^2 z\right) 
	+4 \rho_{0,0} \; m_{0,t}^2 \; z (1-z)^3 \right].
\end{eqnarray}

\subsection{$V \to V H$ Splitting Functions}
\label{subsec:VVH}

The last EW branchings that we need to consider is the case where a Higgs boson radiates from a massive gauge boson. The  helicity amplitudes that correspond to these branchings would be
\begin{equation}
	H_{V\to VH}(z,\tilde{q};\lam_0,\lam_1) = g \; m_0 \sqrt{\frac2{\tilde{q}^2_0-m_0^2}} F_{\lam_0,\lam_1}^{V\to VH},
\end{equation}
with $g = e/\sin \theta_W$ for $V=W^{\pm}$ and $g = e/(\sin \theta_W \cos \theta_W)$ for $V=Z^{0}$. Knowing the polarization vectors of the parent vector bosons, Eqs. (\ref{polT-0}) and (\ref{polLD-0}) and the child vector boson, Eqs. (\ref{eqn:qqguin}) and (\ref{eqn:qqgepsLD}), we can readily calculate the vertex functions,
\begin{equation}
	F_{\lam_0,\lam_1}^{V\to VH} = \sqrt{\frac1{2(\tilde{q}^2_0-m_0^2)}} 
	\left( \epsilon_{\lambda_0} \cdot \epsilon_{\lambda_1}^* \right) .
\end{equation}
The explicit forms of $F_{\lam_0,\lam_1}^{V\to VH}$ functions are given in Table~\ref{tab:VVH}.
\begin{table}
  \begin{center}
  \setlength\extrarowheight{5pt}
    \begin{tabular}{|c|c|c|}
		\hline
		$\lam_0$ & $\lam_1$ & $F_{\lam_0,\lam_1}^{V\to VH}$ \\[5pt]
		\hline \hline 
		+ & + &  $- {m_{0,t} \over \sqrt{2}}$ \\[5pt] 
		\hline
		+ & - & 0 \\[5pt]
		\hline
		+ & $0$ & $\frac{p_{\perp}}{2 \sqrt{\tilde{q}^2 (1-z) z}}$ \\[5pt]
		\hline
		- & + & 0 \\[5pt]
		\hline
		- & - &  $- {m_{0,t} \over \sqrt{2}}$ \\[5pt]
		\hline
		- & $0$ & $-\frac{p_{\perp}}{2 \sqrt{\tilde{q}^2 (1-z) z}}$ \\[5pt]
		\hline
		$0$ & + &  $-\frac{p_{\perp}}{2 z \sqrt{\tilde{q}^2 (1-z) z}}$ \\[5pt]
		\hline
		$0$ & - & $\frac{p_{\perp}}{2 z \sqrt{\tilde{q}^2 (1-z) z}}$ \\[5pt]
		\hline
		$0^*$ & $0^*$ & 0 \\[5pt]
		\hline
    \end{tabular}
  \end{center}
\caption{Spin-unaveraged splitting functions for $V \to V H$. In addition to the factors given above, each term has a phase $e^{i(\lam_0-\lam_1-\lam_2)\phi'}$, where $\lam_{0,1}=\pm,0$ represent the helicity states of the incoming and the outgoing guage vector bosons. The longitudinal polarization states which are marked as $\lam_2=0^*$ give the relevant components in the Dawson's approach \cite{Dawson:1984gx} and can be used to eliminate singularities that appear in the $p_{\perp} \gg m_i, \; i=0,2$ limit.}
\label{tab:VVH}
\end{table}
We can derive different helicity configurations of the $V \to VH$ splitting function as
\begin{eqnarray}
	P_{V\to VH}^{\rm TT}(z,\tilde{q}) &=& \sum_{\lambda_0, \lambda_1= \pm 1} 
	\left| H_{V\to VH}(z,\tilde{q};\lam_0,\lam_1) \right|^2
	\nonumber \\
	&=& {1 \over 2} \; m_{0,t}^2 \; (\rho_{-1,-1}+\rho_{1,1}),
\end{eqnarray}
\begin{eqnarray}
	P_{V\to VH}^{\rm TL}(z,\tilde{q}) &=& 
	\sum_{\lambda_0= \pm 1,  \lambda_1 = 0} 
	\left| H_{V\to VH}(z,\tilde{q};\lam_0,\lam_1) \right|^2
	\nonumber \\
	&=& \frac{1}{4} \; 
	\left[ (1-z) z -m_{0,t}^2 (1-z)^2 - z m_{H,t}^2 \right] \;
	(\rho_{-1,-1}+\rho_{1,1}),
\end{eqnarray}
\begin{eqnarray}
	P_{V\to VH}^{\rm LT}(z,\tilde{q}) &=& 
	\sum_{\lambda_0= 0,  \lambda_1 = \pm 1} 
	\left| H_{V\to VH}(z,\tilde{q};\lam_0,\lam_1) \right|^2
	\nonumber \\
	&=& \frac{1}{2 z^2} \; 
	\left[ (1-z) z -m_{0,t}^2 (1-z)^2 - z m_{H,t}^2 \right] \;
	\rho_{0,0},
\end{eqnarray}
\begin{eqnarray}
	P_{V\to VH}^{\rm LL}(z,\tilde{q}) = 0. 
\end{eqnarray}
Once more, we have used Dawson's approach in the calculation of the LL part to avoid $1/m_i$ terms in massive vector boson splittings. Putting above parts together, the $V\to VH$ splitting function will take the form
\begin{eqnarray}
	P_{V\to VH}(z,\tilde{q}) &=& P_{V\to VH}^{\rm Massless}(z,\tilde{q}) + P_{V\to VH}^{\rm Massive}(z,\tilde{q}),
\end{eqnarray}
with
\begin{eqnarray}
P_{V\to VH}^{\rm Massless}(z,\tilde{q}) &=& \frac{1-z}{4 z}
	\left[ z^2 (\rho_{-1,-1} + \rho_{1,1}) + 2 \rho_{0,0} \right],
\nonumber \\
P_{V\to VH}^{\rm Massive}(z,\tilde{q}) &=& -\frac{m_{H,t}^2 }{4 z}
	\left[ z^2 (\rho_{-1,-1} + \rho_{1,1}) + 2 \rho_{0,0} \right]
	\nonumber \\
	&-& 
	\frac{m_{0,t}^2 }{4 z^2} 
	\left[ \left( 2 z^2-4 z+2\right) \rho_{0,0}+\left(z^4-2 z^3-z^2 \right) 
	(\rho_{-1,-1}+\rho_{1,1}) \right].
\end{eqnarray}
Having calculated all relevant EW splitting functions, we are now in a position to implement our EW parton shower in \textsf{Herwig 7}. A brief note on the changes in \textsf{Herwig 7} interface can be found in Appendix \ref{sec:HI}, which includes the newly introduced commands. In the next section, we will present our results for testing this implementation against fixed-order (FO) EW radiations. We will also test the performance of this EW parton shower in the prediction of high-energy scattering events.

\section{Results and Discussions}
\label{sec:res}

To test the performance of our EW parton shower, we choose a number of events that can be meaningfully showered with specific EW branching classes in \textsf{Herwig 7}. Then, firstly, we generate and shower these events using an EW-only parton shower scheme while limiting the parton shower to just one FS emission. This would allow us to collect the corresponding single-step EW resummation data. Secondly, we calculate the equivalent FO contributions, using the relevant hard MEs and without EW shower. Comparing these results would produce a good performance test for the implemented parton shower.  

\begin{figure}
\centering
\includegraphics[width=.6\textwidth]{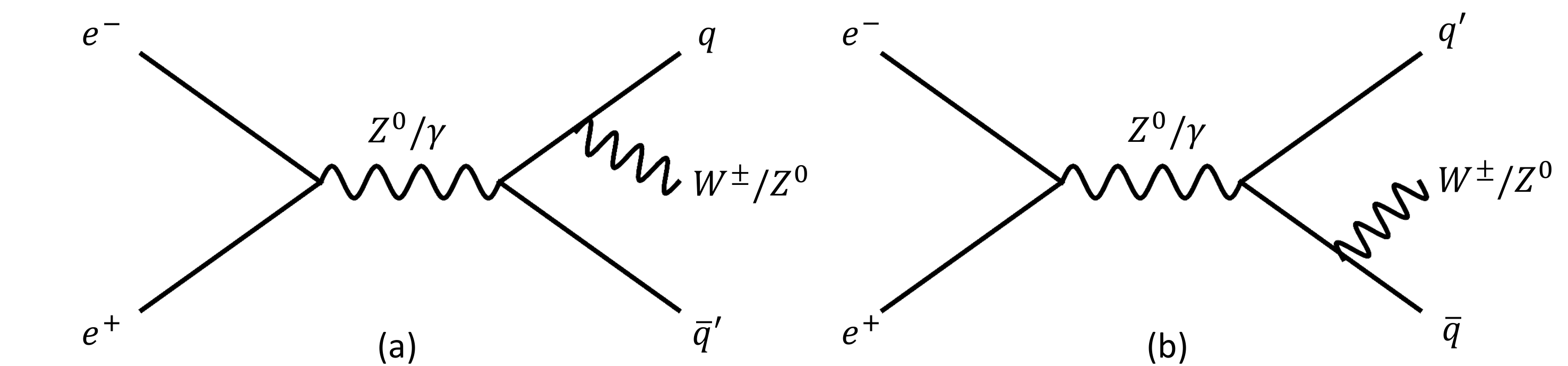}
\caption{Fixed order equivalent channels for showering a $e^- e^+ \to Z^0/\gamma \to q \bar{q}$ process with a single-step $q \to q' V$ branching.}
\label{FO-qqV-diags}
\end{figure}

Hence, for the case of $q \to q' V$ splittings, we choose the $e^- e^+ \to Z^0/\gamma \to q \bar{q}$ process to be the source of resummed contributions from the EW shower. The equivalent FO calculation can be carried out using the $e^- e^+ \to Z^0 \to q \bar{q}' V$ channels, shown in Figure~\ref{FO-qqV-diags}. The corresponding MEs for the FO calculations are generated by \textsf{MadGraph5} \cite{Alwall:2014hca} while for the resummed computations we use a \textsf{Herwig 7} internal ME, \texttt{MEee2gZ2qq} \cite{Platzer:2011bc,Bellm:2019wrh,Bellm:2020}. The produced events are analysed by \textsf{Rivet} \cite{Buckley:2010ar}. 

\begin{figure}
\begin{subfigure}{.5\textwidth}
  \centering
  \includegraphics[width=1\linewidth]{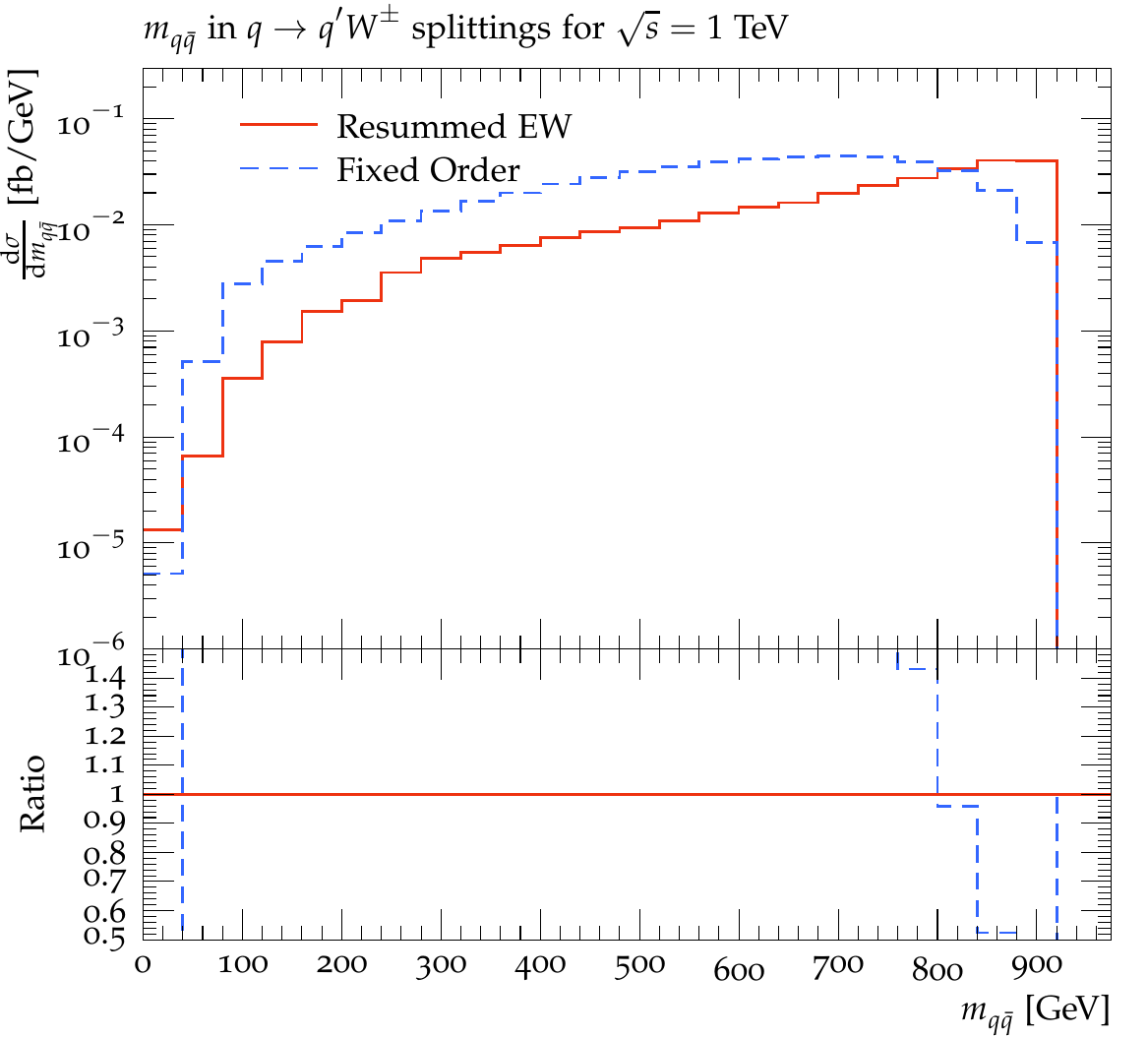}
  \caption{}
  \label{}
\end{subfigure}%
\begin{subfigure}{.5\textwidth}
  \centering
  \includegraphics[width=1\linewidth]{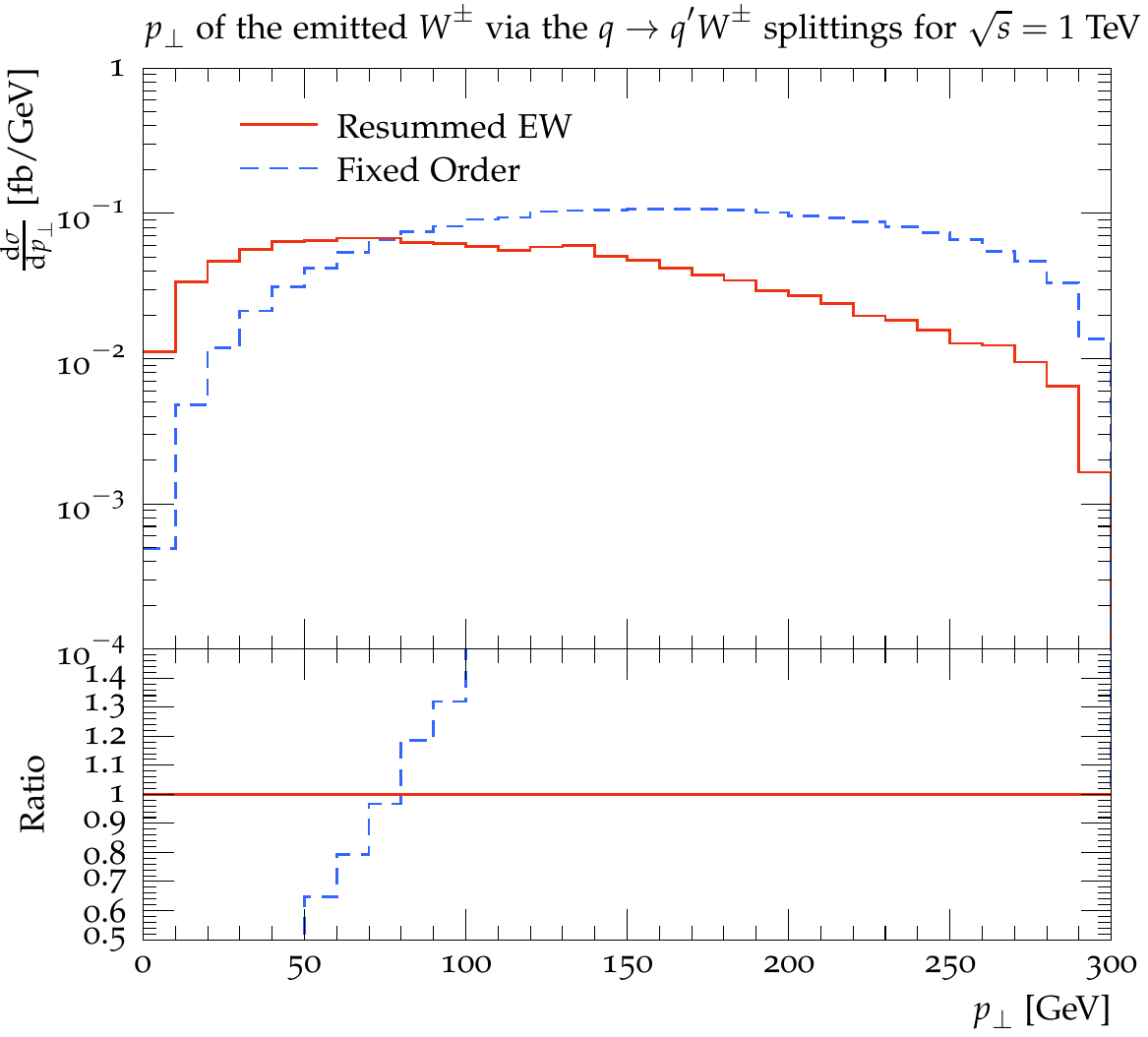}
  \caption{} 
  \label{}
\end{subfigure}
\begin{subfigure}{.5\textwidth}
  \centering \hspace{0.1in}
  \textcolor{white}{...}\includegraphics[width=1\linewidth]{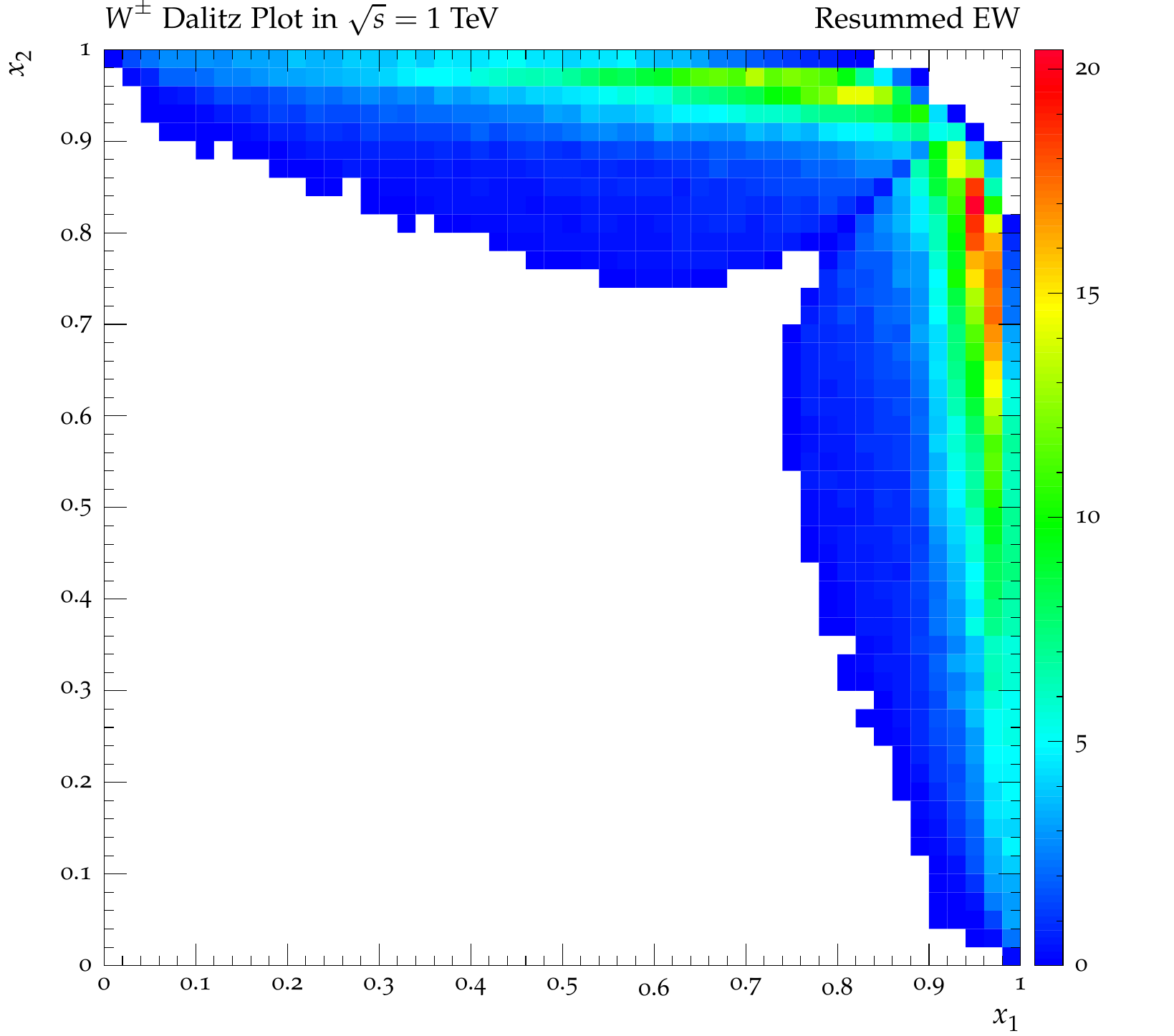}
  \caption{}
  \label{}
\end{subfigure}%
\begin{subfigure}{.5\textwidth}
  \centering
  \textcolor{white}{...}\includegraphics[width=1\linewidth]{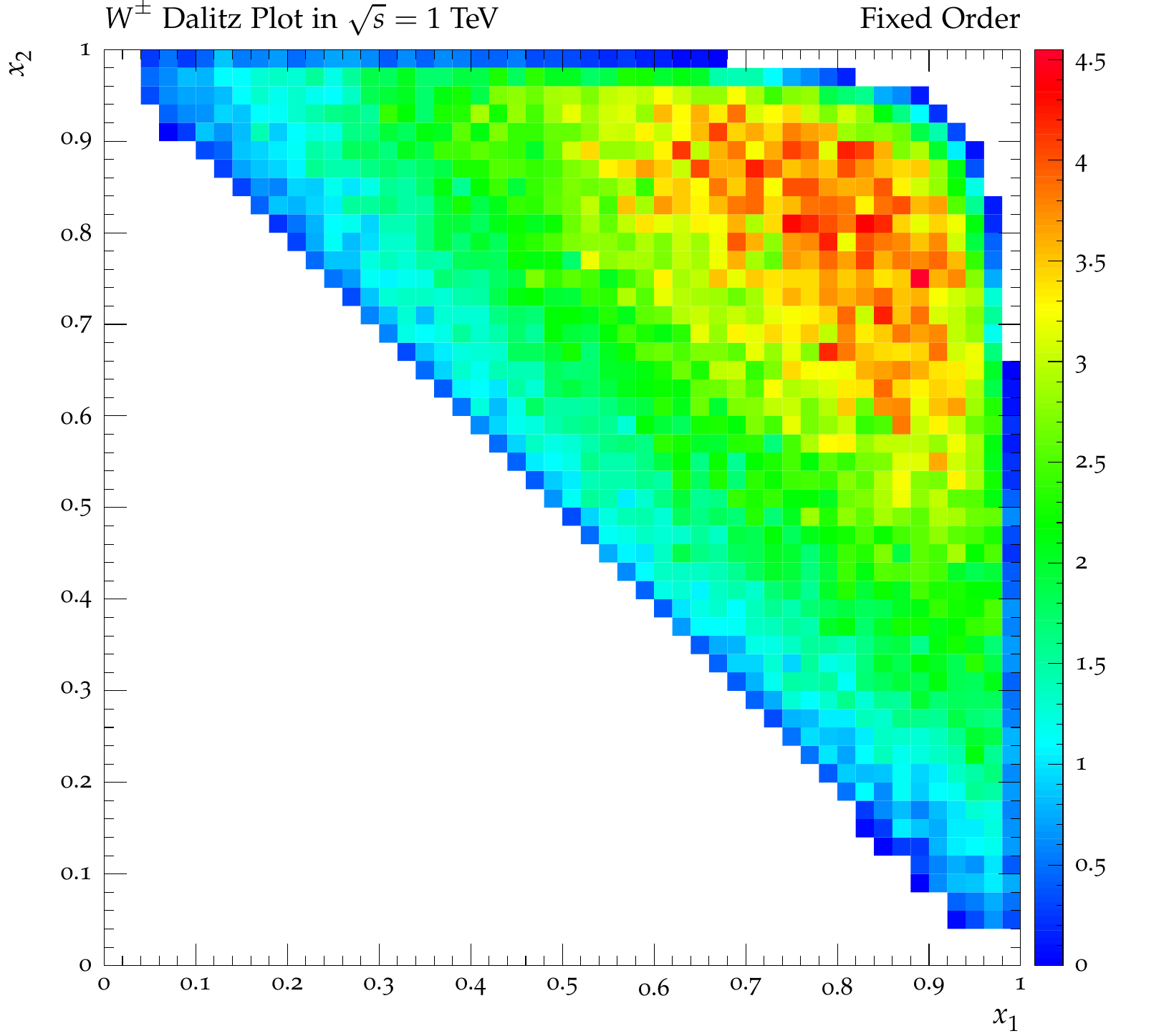}
  \caption{}
  \label{}
\end{subfigure}
\caption{Performance test for $q \to q' W^{\pm}$ EW branching in \textsf{Herwig 7} for $\sqrt{s}=1$ TeV. (a) and (b) show the differential rate of $W^{\pm}$ emissions respectively as functions of the mass of the quark-antiquark system, $m_{q\bar{q}}$ and the transverse momenta of radiated gauge bosons, $p_{\perp}$. (c) and (d) illustrate the differential rate of $W^{\pm}$ emissions as functions of the light-cone fraction of the momentum of the quark-antiquark pairs, $x_1$ and $x_2$, in resummed EW and FO calculations, respectively.}
\label{qqW-1TeV}
\end{figure}

\begin{figure}
\begin{subfigure}{.5\textwidth}
  \centering
  \includegraphics[width=1\linewidth]{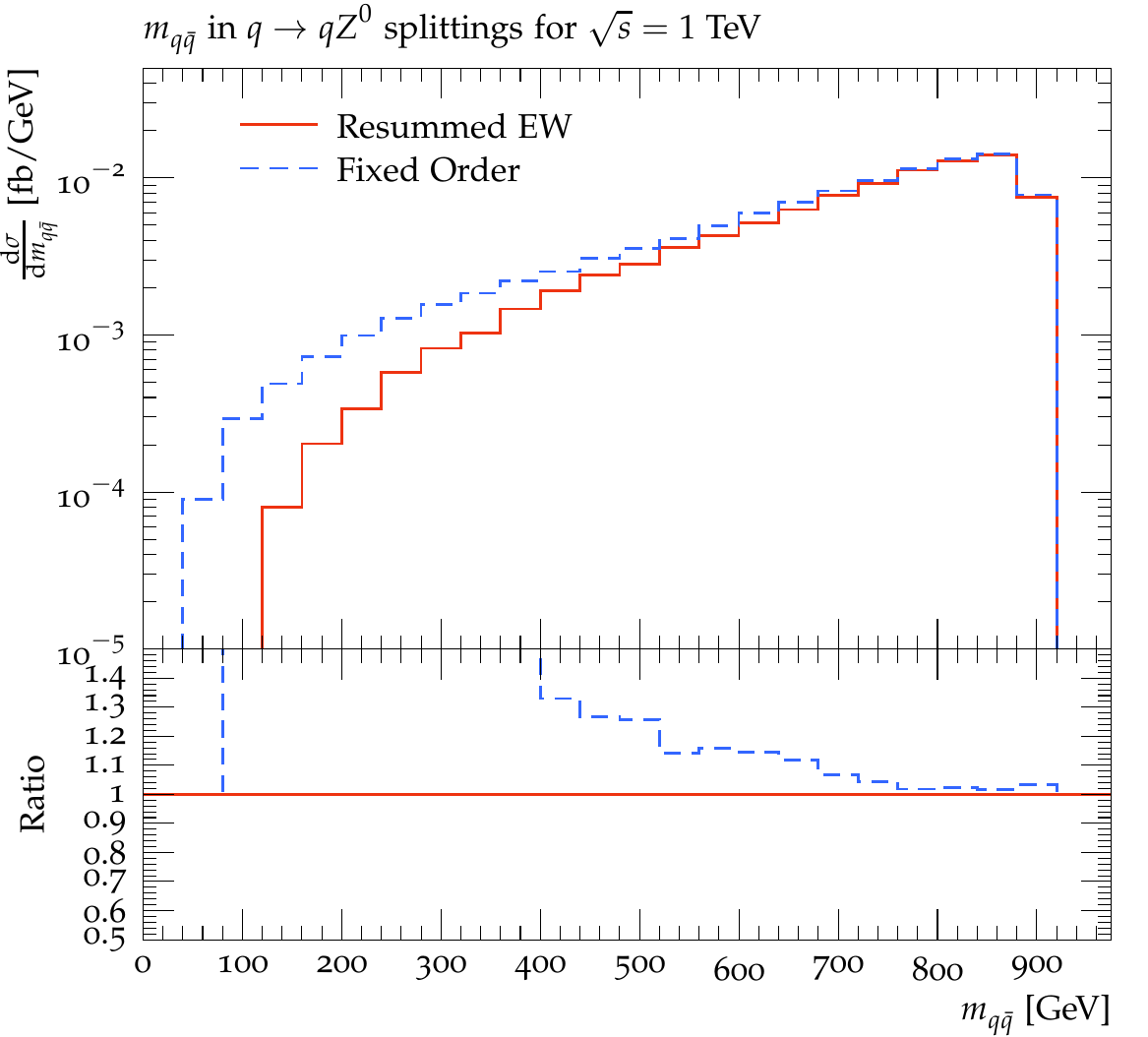}
  \caption{}
  \label{}
\end{subfigure}%
\begin{subfigure}{.5\textwidth}
  \centering
  \includegraphics[width=1\linewidth]{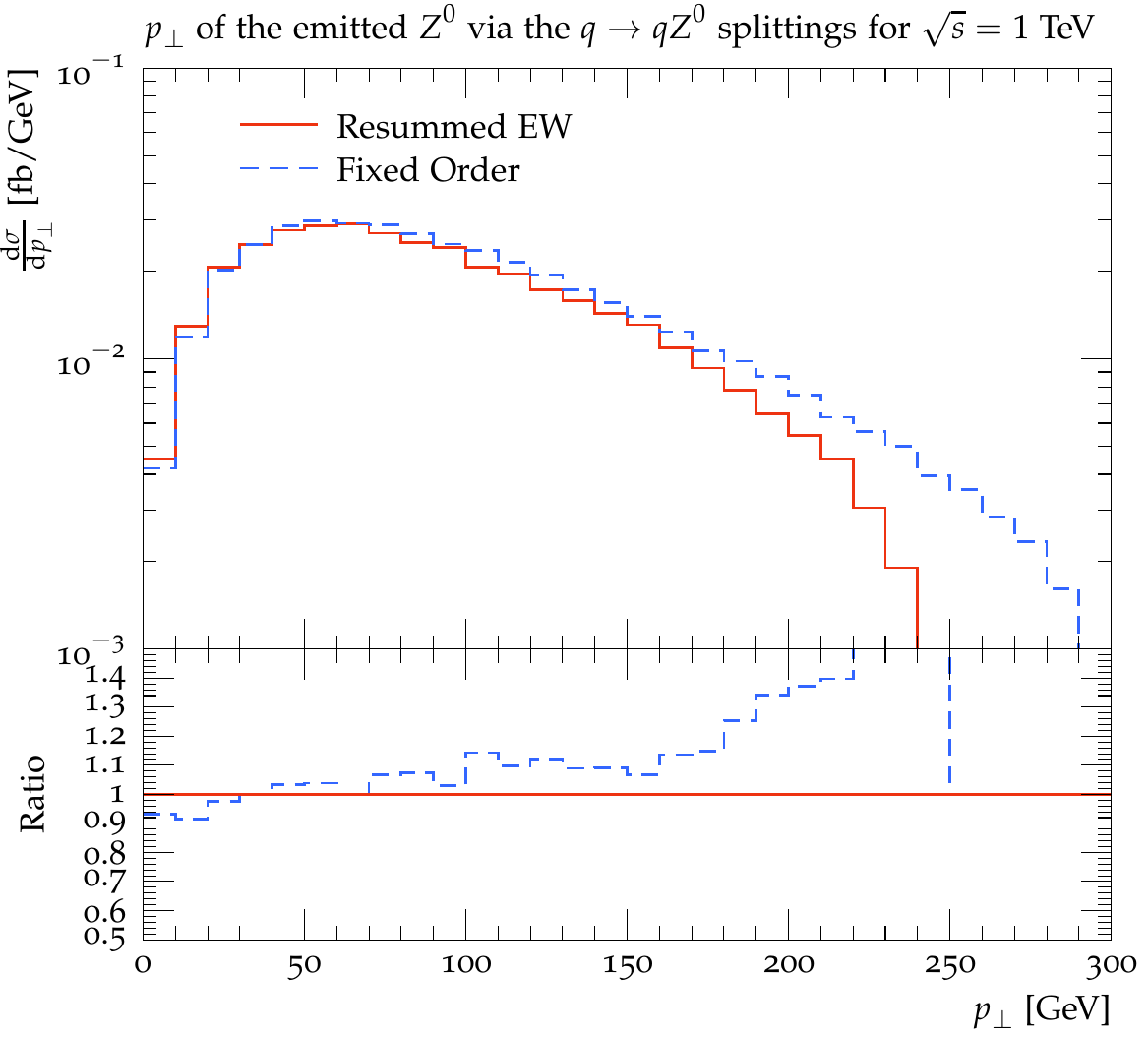}
  \caption{} 
  \label{}
\end{subfigure}
\begin{subfigure}{.5\textwidth}
  \centering \hspace{0.1in}
  \textcolor{white}{...}\includegraphics[width=1\linewidth]{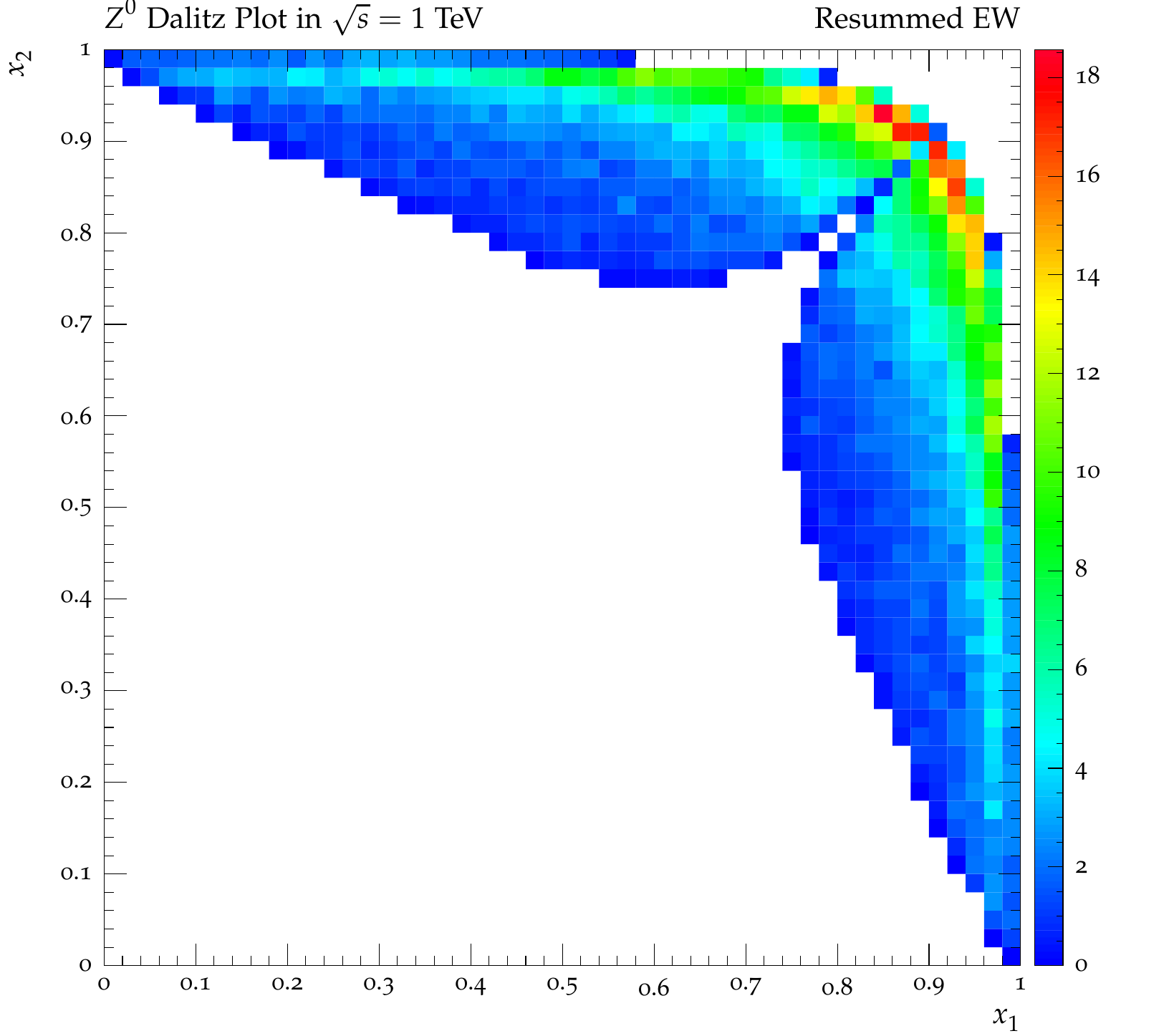}
  \caption{}
  \label{}
\end{subfigure}%
\begin{subfigure}{.5\textwidth}
  \centering
  \textcolor{white}{...}\includegraphics[width=1\linewidth]{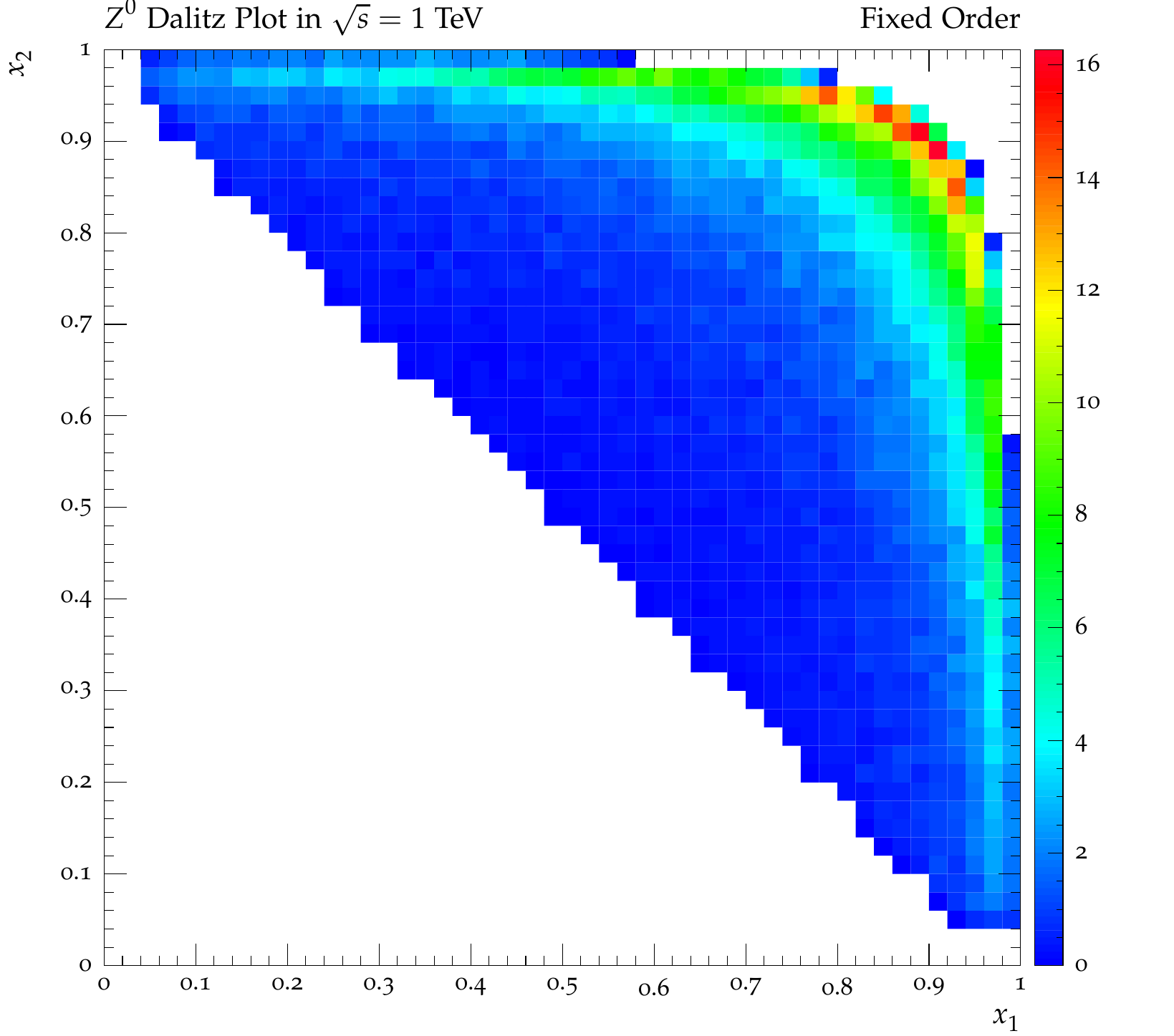}
  \caption{}
  \label{}
\end{subfigure}
\caption{Performance test for $q \to q Z^0$ EW branching in \textsf{Herwig 7} for $\sqrt{s}=1$ TeV. The notation of the figure is the same as in Figure~\ref{qqW-1TeV}.}
\label{qqZ-1TeV}
\end{figure}

Figures~\ref{qqW-1TeV} and~\ref{qqZ-1TeV} show the results of this analysis, respectively for the $q \to q' W^{\pm}$ and $q \to q Z^0$ splitting functions in $\sqrt{s}=1$ TeV center-of-mass energy. The panels (a) and (b) of these figures demonstrate the differential rates of the EW gauge boson emissions as functions of the mass of the quark-antiquark systems, $m_{q\bar{q}}$, and the transverse momenta of the radiated gauge bosons, $p_{\perp}$. The red solid histograms show the kinematics of the gauge bosons that have been emitted as the results of the implemented EW parton shower, up to a single emission, and hence labelled \textit{EW Resummed}. These are compared against their FO counterparts, presented as blue dashed histograms. Furthermore, the Dalitz plots in (c) panels of these figures show the normalized weights of the gauge boson PS radiations as functions of the light-cone momentum fractions of the final-state quarks. These results can be compared against similar FO plots in panels (d). 

\begin{figure}
\begin{subfigure}{.5\textwidth}
  \centering
  \includegraphics[width=1\linewidth]{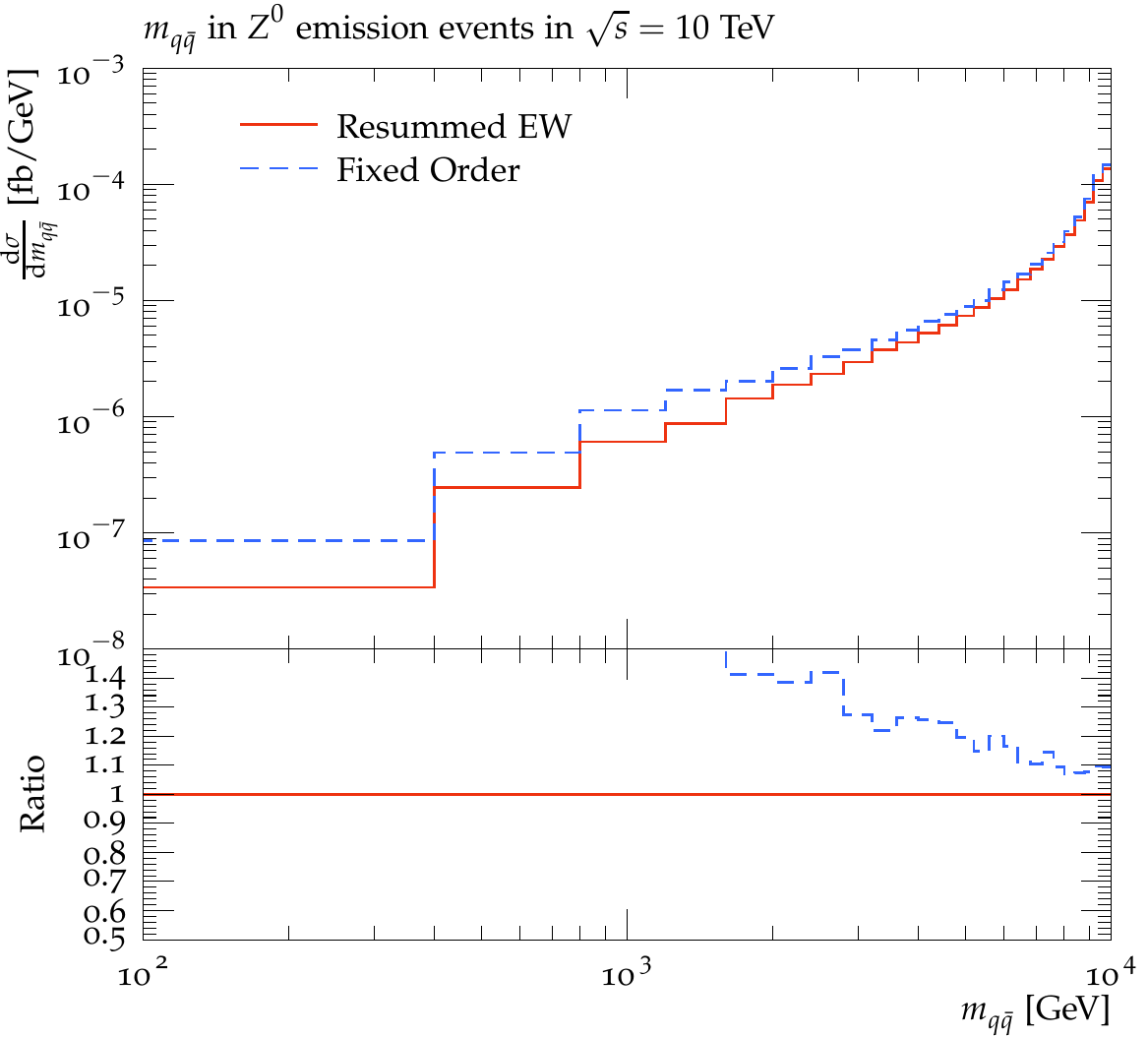}
  \caption{}
  \label{}
\end{subfigure}%
\begin{subfigure}{.5\textwidth}
  \centering
  \includegraphics[width=1\linewidth]{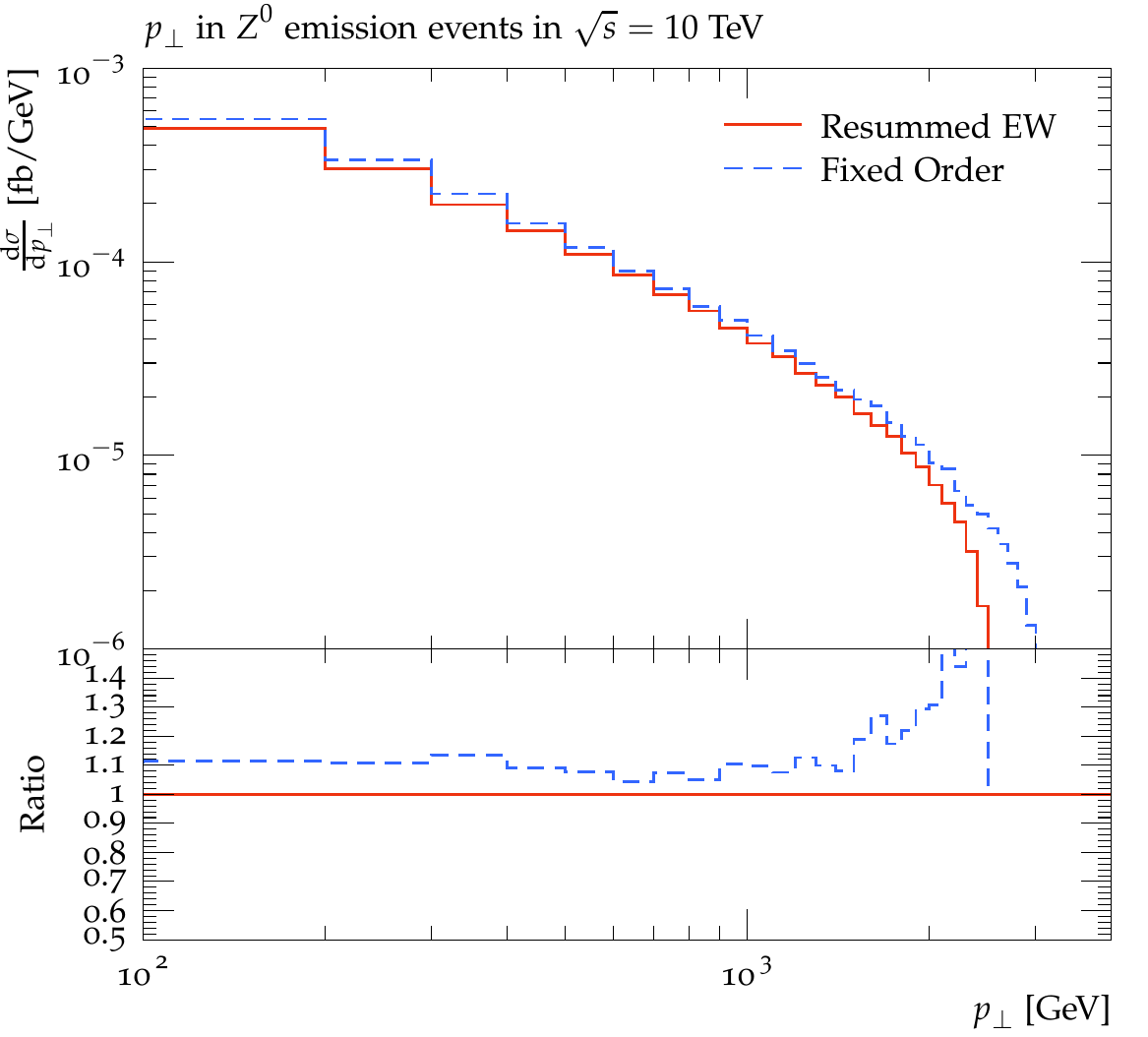}
  \caption{} 
  \label{}
\end{subfigure}
\begin{subfigure}{.5\textwidth}
  \centering \hspace{0.1in}
  \textcolor{white}{...}\includegraphics[width=1\linewidth]{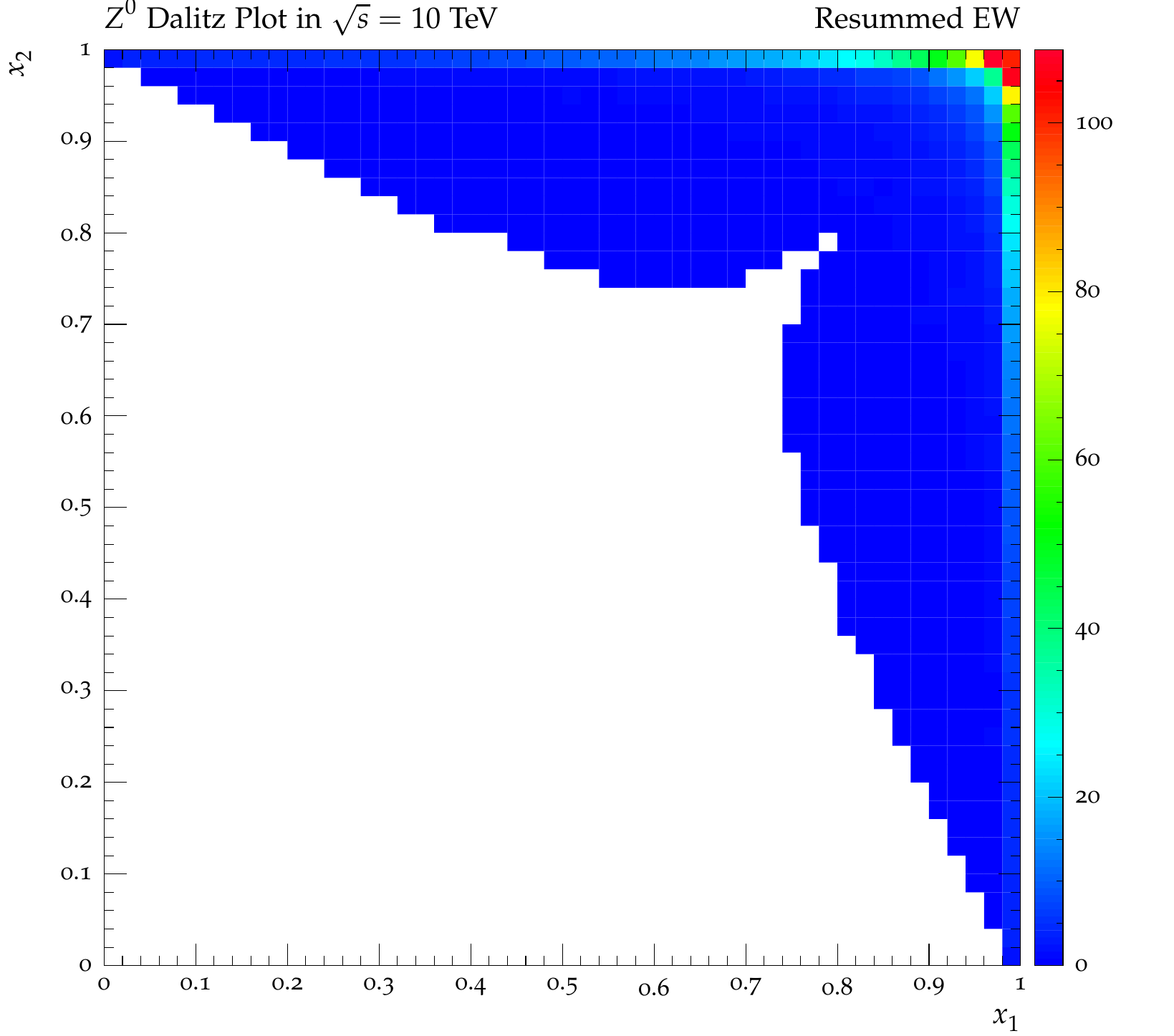}
  \caption{}
  \label{}
\end{subfigure}%
\begin{subfigure}{.5\textwidth}
  \centering
  \textcolor{white}{...}\includegraphics[width=1\linewidth]{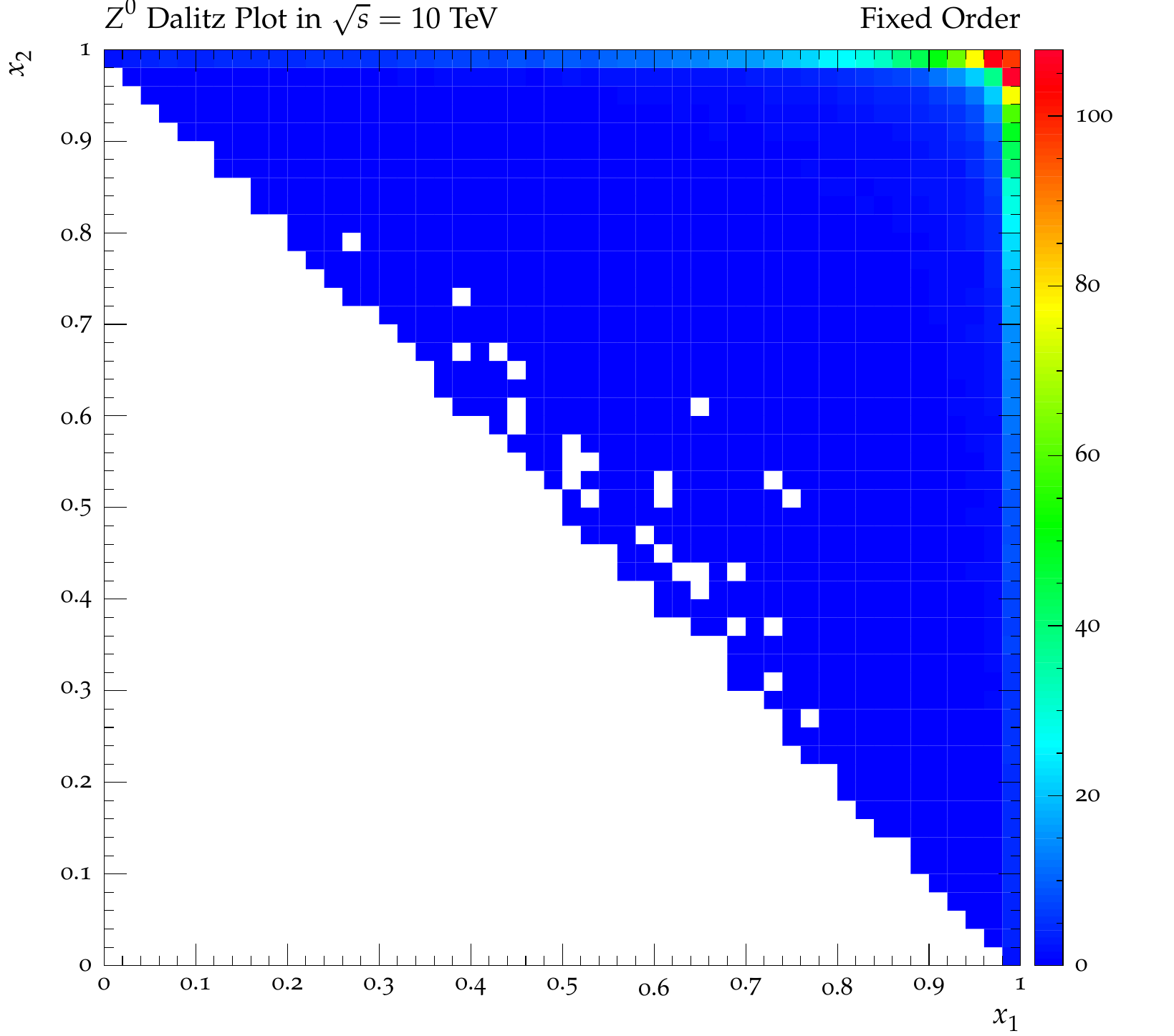}
  \caption{}
  \label{}
\end{subfigure}
\caption{Performance test for $q \to q Z^0$ EW branching in \textsf{Herwig 7} for $\sqrt{s}=10$ TeV. The notation of the figure is the same as in Figure~\ref{qqZ-1TeV}.}
\label{qqZ-10TeV}
\end{figure}

One can immediately recognize that the single-step EW radiations from the $q \to q Z^0$ splittings have a relatively good agreement with their FO counterparts and the observed discrepancies in the small mass region of the $q\bar{q}$-pair system, or in the high-$p_{\perp}$ tail of the radiated gauge bosons, are mainly the remnants of the collinear factorisation approximation, the Eq. (\ref{collaprox}). This conclusion is particularly fortified by observing the same pattern in Figure~\ref{qqZ-10TeV}, where a similar calculation has been made with $\sqrt{s}=10$ TeV. In this latter case, the agreement between the EW resummed and the FO results has been expectedly improved, since the collinear factorisation theorem produces better results with increasing the factorisation scale. Moreover, the performance test for the $q \to q' W^{\pm}$ EW branching in Figure~\ref{qqW-1TeV}, although producing comparable results, shows slightly different behaviours compared to the FO calculations. This is due to the difference between the treatment of \textsf{MadGraph5} towards the longitudinal components of the $q \to q' W^{\pm}$ splitting, and our use of Dawson's approach. These differences may become significant for high-energy FO results but would be of no consequence for successive radiations in an EW parton shower scheme. Additionally, the \textit{wing} shapes of the Dalitz plots are produced by symmetric radiations of the gauge bosons from the quark-antiquark pairs, showing the statical balance of the EW radiations.

\begin{figure}
\begin{subfigure}{.5\textwidth}
  \centering
  \includegraphics[width=1\linewidth]{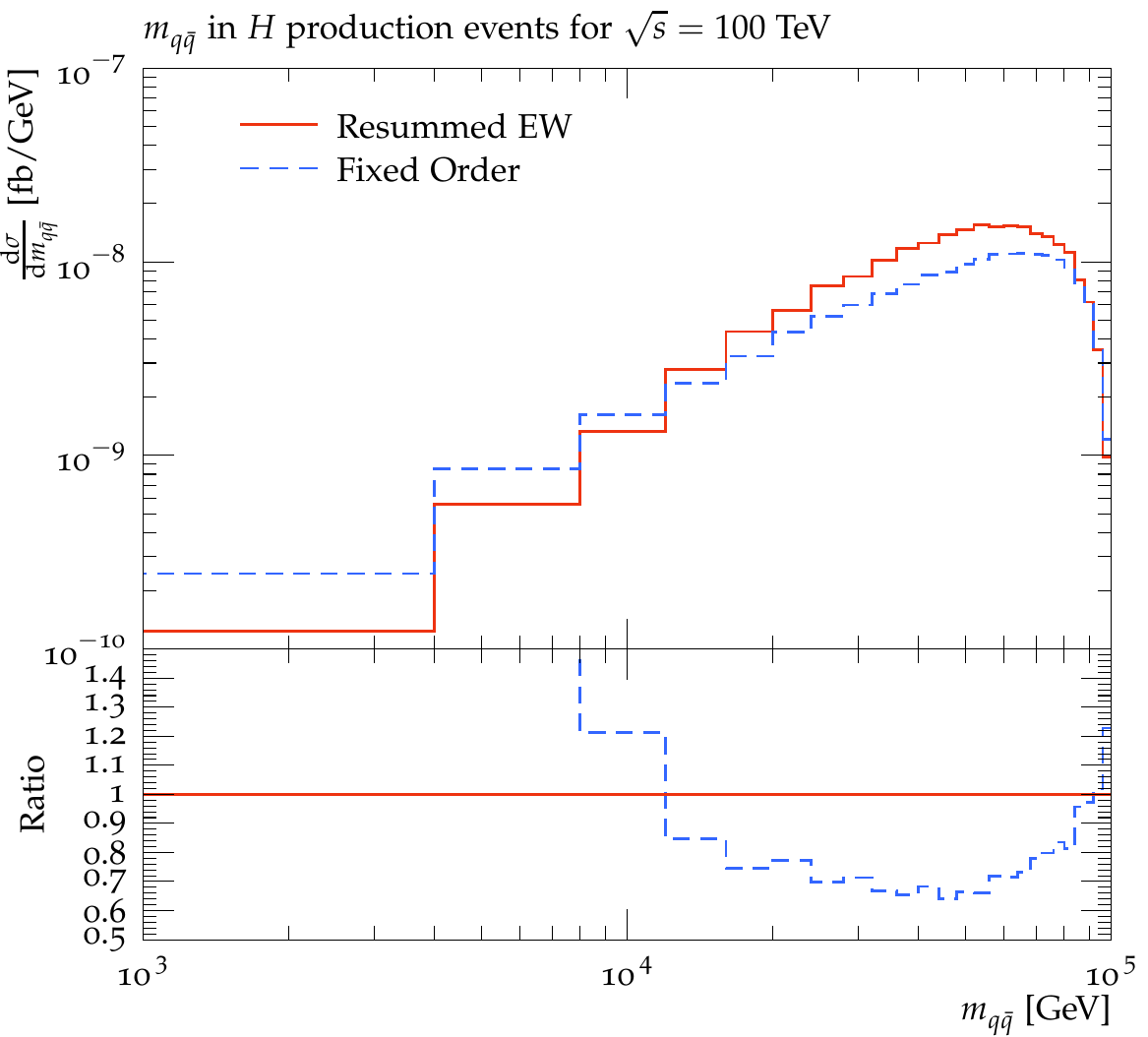}
  \caption{}
  \label{}
\end{subfigure}%
\begin{subfigure}{.5\textwidth}
  \centering
  \includegraphics[width=1\linewidth]{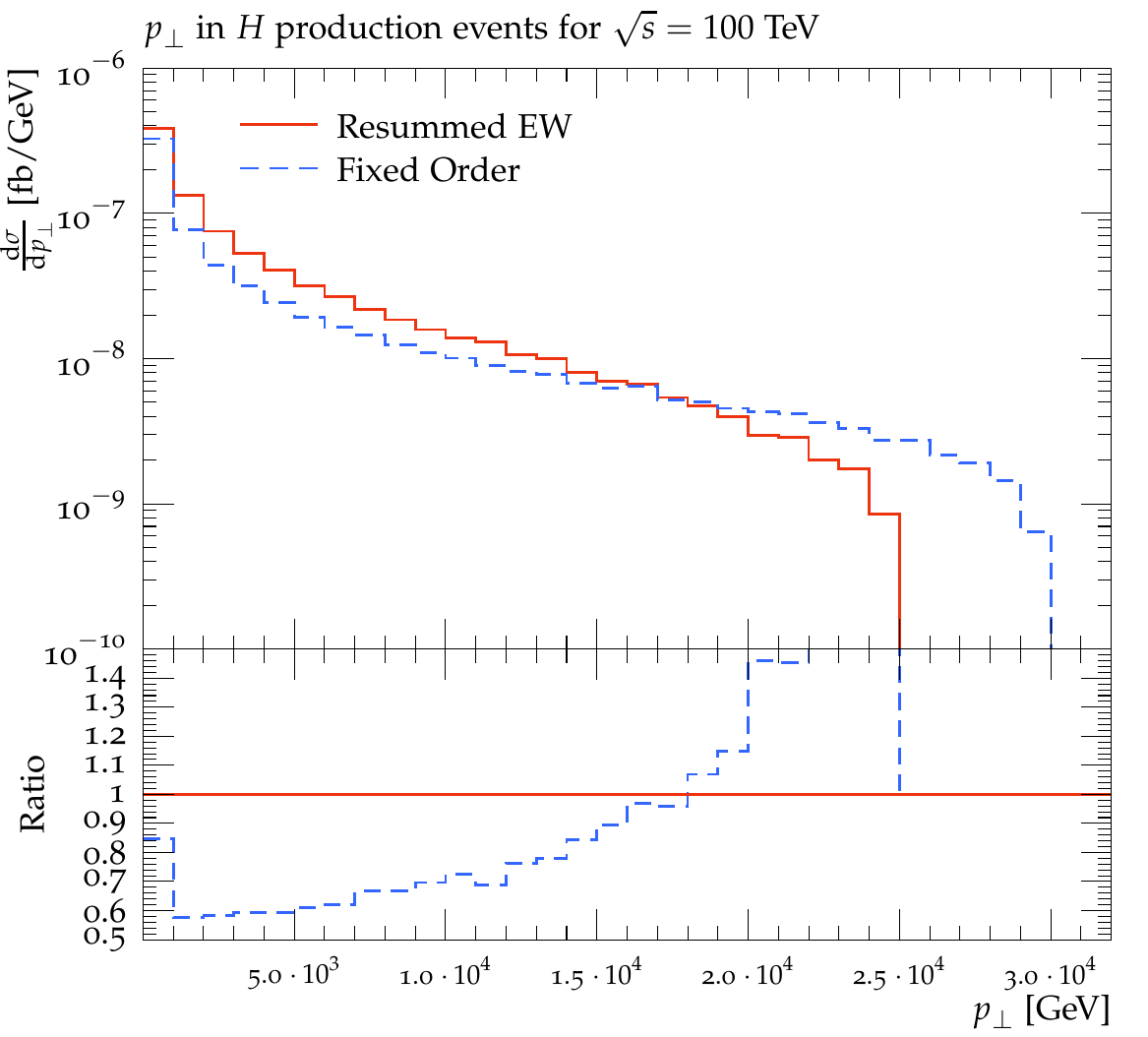}
  \caption{} 
  \label{}
\end{subfigure}
\begin{subfigure}{.5\textwidth}
  \centering \hspace{0.1in}
  \textcolor{white}{...}\includegraphics[width=1\linewidth]{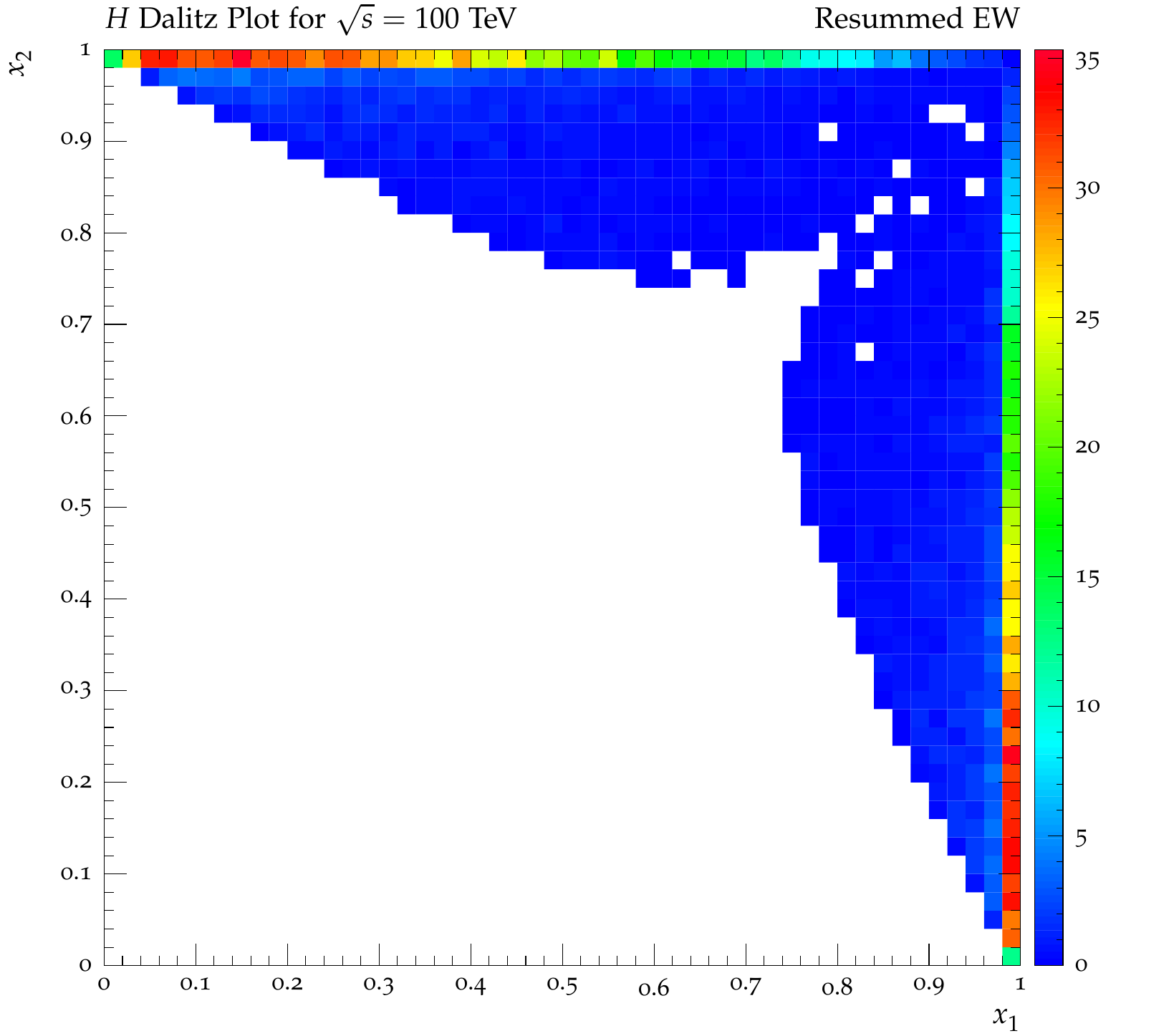}
  \caption{}
  \label{}
\end{subfigure}%
\begin{subfigure}{.5\textwidth}
  \centering
  \textcolor{white}{...}\includegraphics[width=1\linewidth]{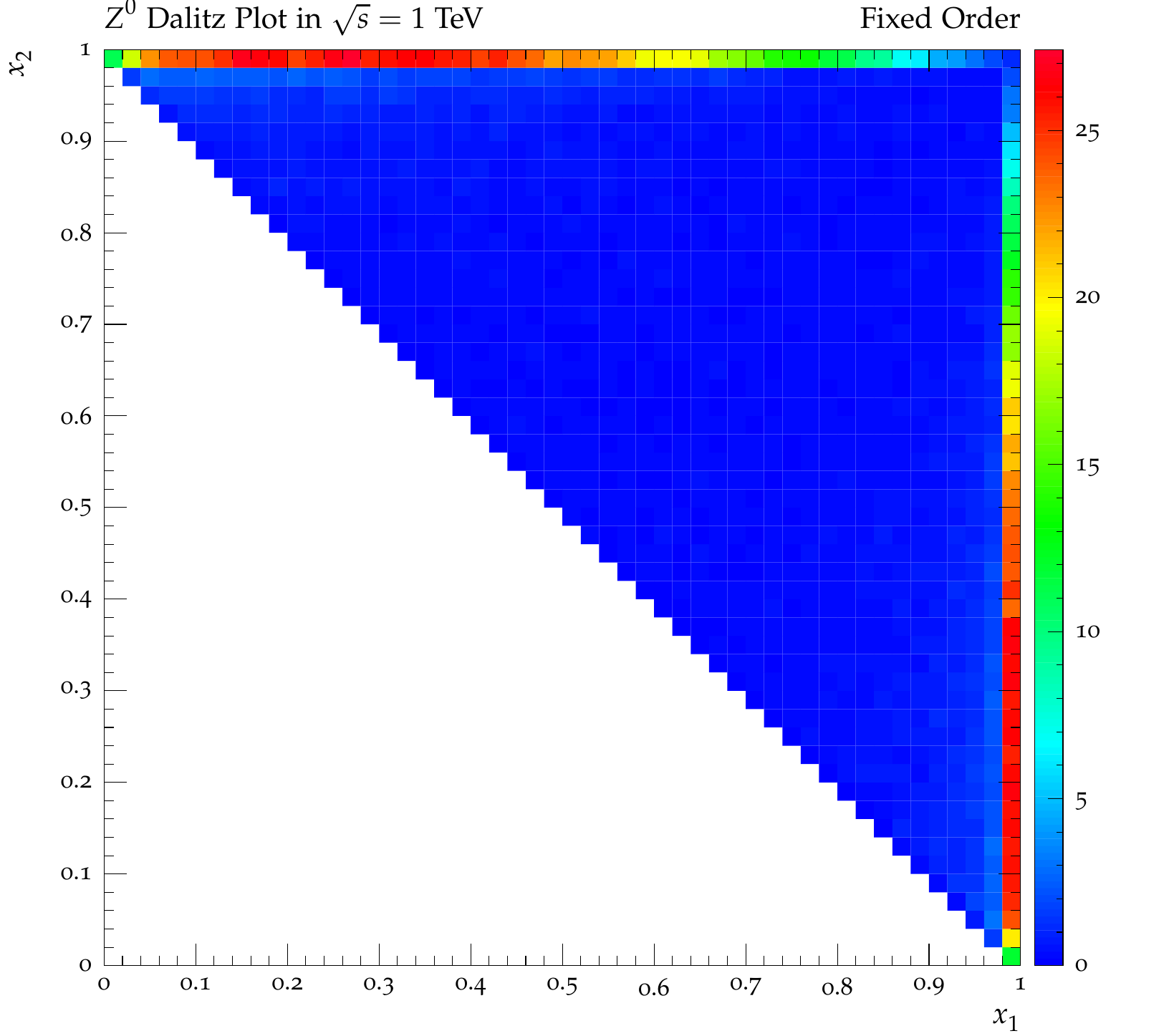}
  \caption{}
  \label{}
\end{subfigure}
\caption{Performance test for $q \to q' Z^0$ EW branching in \textsf{Herwig 7} for $\sqrt{s}=10$ TeV. The notation of the figure is the same as in Figure~\ref{qqW-1TeV}.}
\label{qqH-100TeV}
\end{figure}

We can also perform a very similar test for $q \to q H$ breaching, using the same $e^- e^+ \to Z^0/\gamma \to q \bar{q}$ process. This time, we collect the appropriate EW resummation data for $\sqrt{s}=100$ TeV, ensuring that the energy scale can go high enough to allow for high-energy splittings \textit{i.e.} $t \to tH$ and $b \to bH$. Furthermore, the equivalent FO calculations can be done in similar channels as shown in Figure~\ref{FO-qqV-diags}, by replacing the final state gauge bosons with Higgs bosons. Again, the corresponding FO MEs are produced by \textsf{MadGraph5} while the internal \texttt{MEee2gZ2qq} scattering amplitude is showered by EW radiations, resulting in the single-step resummed EW data. These results are plotted and compared in Figure~\ref{qqH-100TeV}, with the same general layout as in Figure~\ref{qqW-1TeV}. It can be seen that the resummed calculations produce a relatively good description of the FO data. 

\begin{figure}
\centering
\includegraphics[width=.9\textwidth]{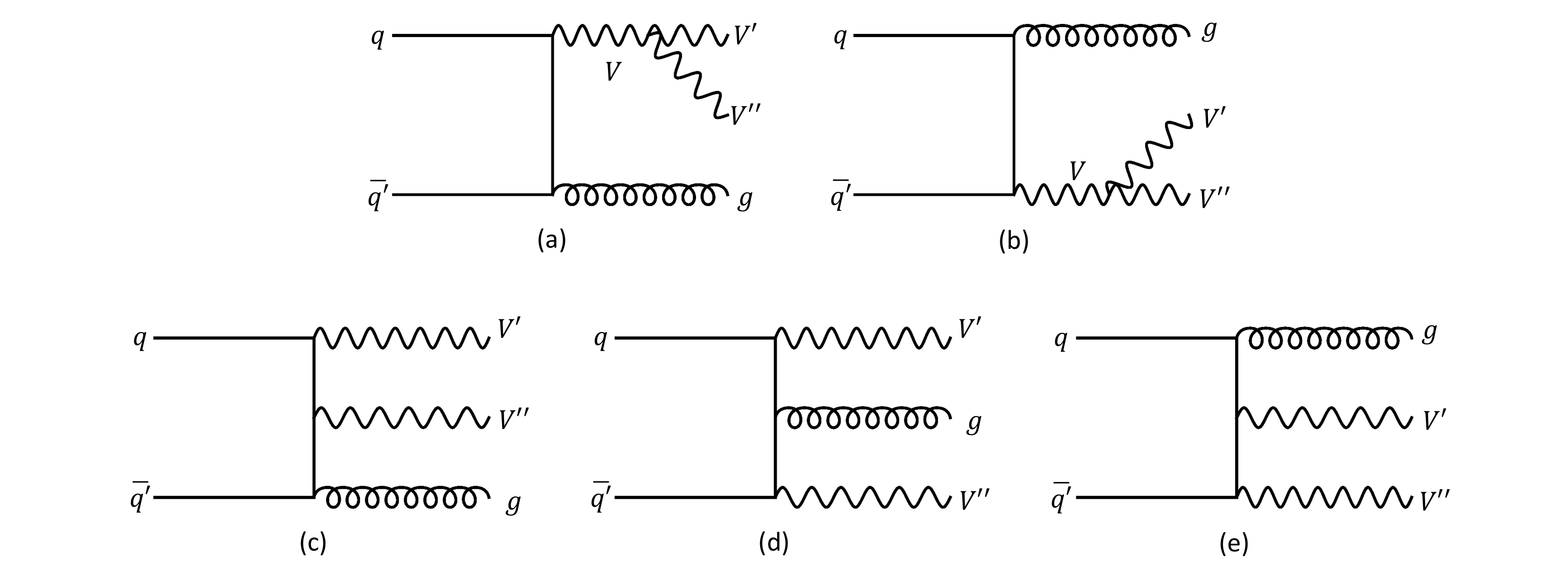}
\caption{Diagrams (a) and (b) are the Fixed order equivalent channels for showering a $q \bar{q} \to V + Jet$ process with a single-step $V\to V' V''$ branching. For the cases involving $W^+W^-$ pairs in the final-state, diagrams (c), (d) and (e) must also be included to preserve gauge invariance. However, the contributions of these channels can be suppressed by introducing angular separation cuts $\Delta R_{W^+,W^-} > 1.0$ and $\Delta R_{W^{\pm},Jet} < 1.0$. Additionally, to ensure a clean branching signature, we impose a $k_{\perp}^{\rm jet}>1$ TeV cut on the transverse momentum of the produced jets.}
\label{FO-VVV-diags}
\end{figure}

To check the performances of the remaining splitting classes, \textit{i.e.} $V \to V' V''$ and $V \to V H$, we choose a $q \bar{q} \to Vg$ underlying event with $q=u,\; d$ at $\sqrt{s}=13$ TeV energy scale and shower it with EW radiations to produce the EW resummation data. The corresponding FO calculations can also be done, using a set of production channels that are given in Figure~\ref{FO-VVV-diags}. One should note that in the case of a $V \to V H$ branching, only the diagrams (a) and (b) of Figure~\ref{FO-VVV-diags} (with replacing one of the FS gauge bosons with a Higgs boson) are needed in the FO calculations. For the $V \to V' V''$ cases, diagrams (c), (d) and (e), although being irrelevant to the EW radiations, must be included in the FO calculations to preserve the gauge invariance. This would also mean that we cannot directly separate $Z^0 \to W^+ W^-$ and $\gamma \to W^+ W^-$ radiations from each other and from the contributions of the (c), (d) and (e) channels. We will, however, include angular separation cuts in our FS analysis to suppress the unwanted contributions. Additionally, to ensure a clean EW branching signature, we impose a $k_{\perp}^{\rm jet}>1$ TeV cut on the transverse momentum of the produced jets on both the EW resummed and the FO calculations. 

\begin{figure}
\begin{subfigure}{.5\textwidth}
  \centering
  \includegraphics[width=1\linewidth]{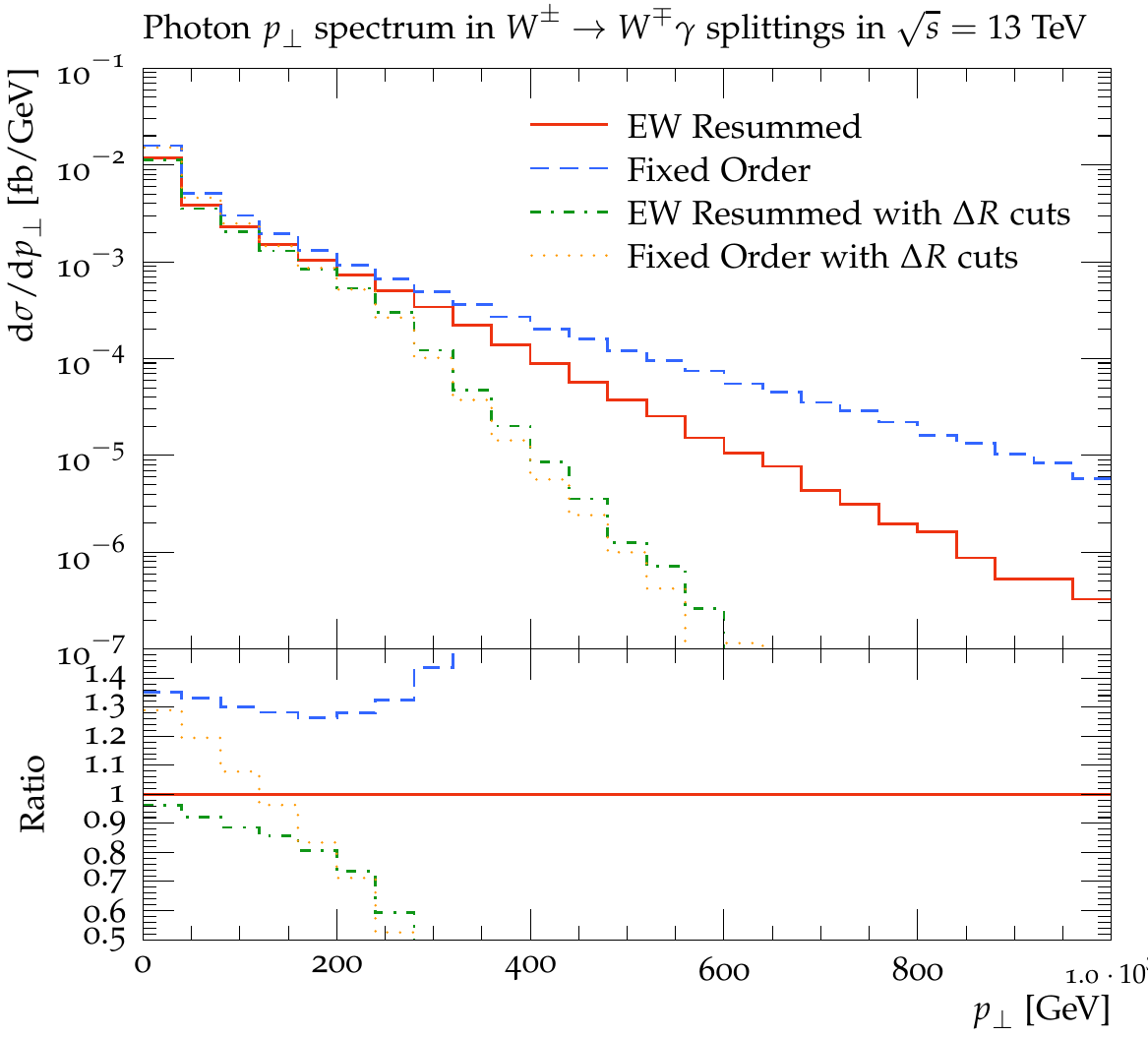}
  \caption{}
  \label{}
\end{subfigure}%
\begin{subfigure}{.5\textwidth}
  \centering
  \includegraphics[width=1\linewidth]{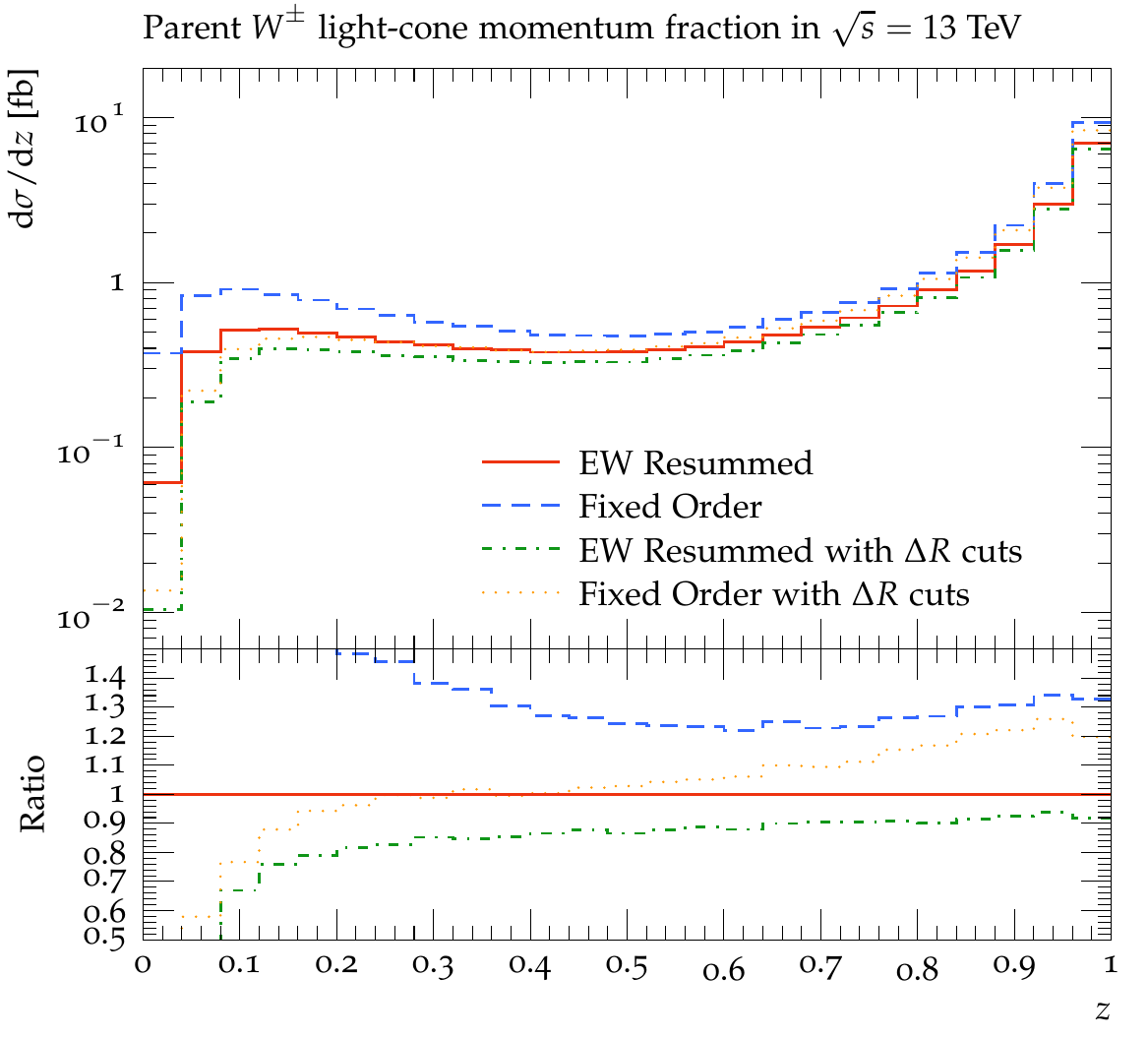}
  \caption{} 
  \label{}
\end{subfigure}
\caption{Performance test for $W^{\pm} \to W^{\pm} \gamma$ EW branching in \textsf{Herwig 7} for $\sqrt{s}=13$ TeV. The notation of the plot is the same as in Figure~\ref{wwz-13TeV}.}
\label{wwg-13TeV}
\end{figure}

Therefore, to test the performance of the $W^{\pm} \to W^{\pm} V$ EW branchings, with $V=Z^0, \; \gamma$, we use a $q \bar{q} \to W^{\pm} g$ underlying event that is showered with the corresponding EW radiations. The results are presented in Figures~\ref{wwz-13TeV} and \ref{wwg-13TeV}, respectively. In each of these figures, panel (a) shows the differential production rate of the emitted gauge boson as a function of its transverse momentum while panel (b) demonstrates the distribution of the light-cone momentum fraction of the parent gauge boson. In both cases, the EW resummed results are plotted as red solid histograms while the FO calculations are the blue dashed histograms. It can be observed that the EW resummed results behave similarly to the FO calculations and seem able to produce a reasonably sound description of the latter. Meanwhile, the discrepancy in the large-$p_{\perp}$ tail is a direct consequence of the extra contributions, coming from the channels that are shown in diagrams (c), (d) and (e) of Figure~\ref{FO-VVV-diags}. One way of suppressing these contributions is to impose a set of angular separation cuts between the FS particles and the gluon-tagged jet, say
\begin{equation}
\Delta R_{W^{\pm},V} > 1, \qquad \Delta R_{W^{\pm},jet} < 1, \qquad \Delta R_{V,jet}< 1.
\label{DRcuts}
\end{equation}
We have included these cuts on our analysis and plotted the corresponding results with green dash-dotted and orange dotted histograms for the EW single-step resummation and FO calculations, respectively. It is immediately apparent that with the irrelevant contributions suppressed, the resummed calculations have a much better agreement with the FO results. One should also note that the closeness of the EW resummed and the FO results in the panels (b) show that the kinematic variables of the parent gauge bosons (in the EW shower) and the exchanged gauge boson (in the FO events) are nearly identical.

\begin{figure}
\begin{subfigure}{.5\textwidth}
  \centering
  \includegraphics[width=1\linewidth]{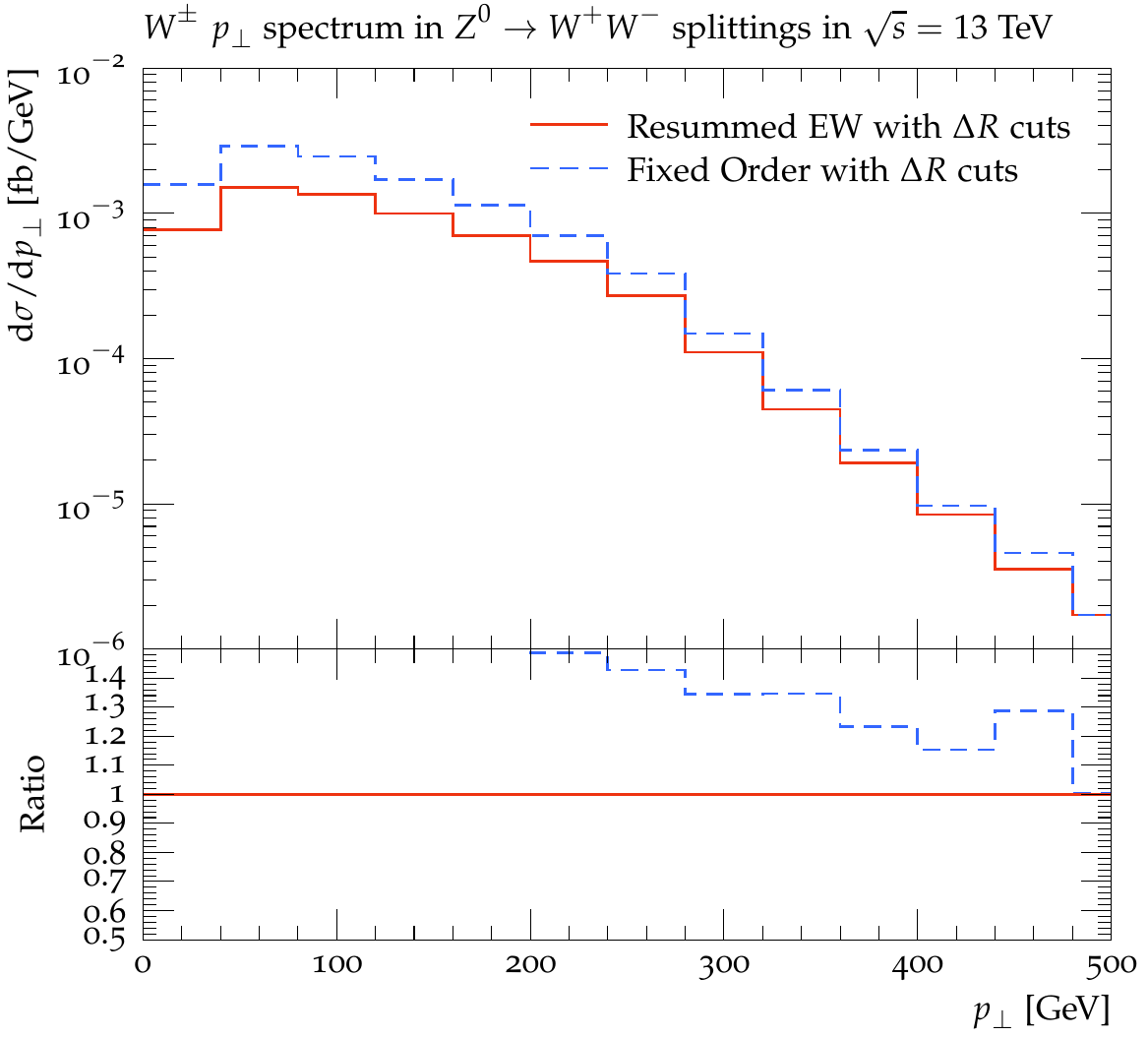}
  \caption{}
  \label{}
\end{subfigure}%
\begin{subfigure}{.5\textwidth}
  \centering
  \includegraphics[width=1\linewidth]{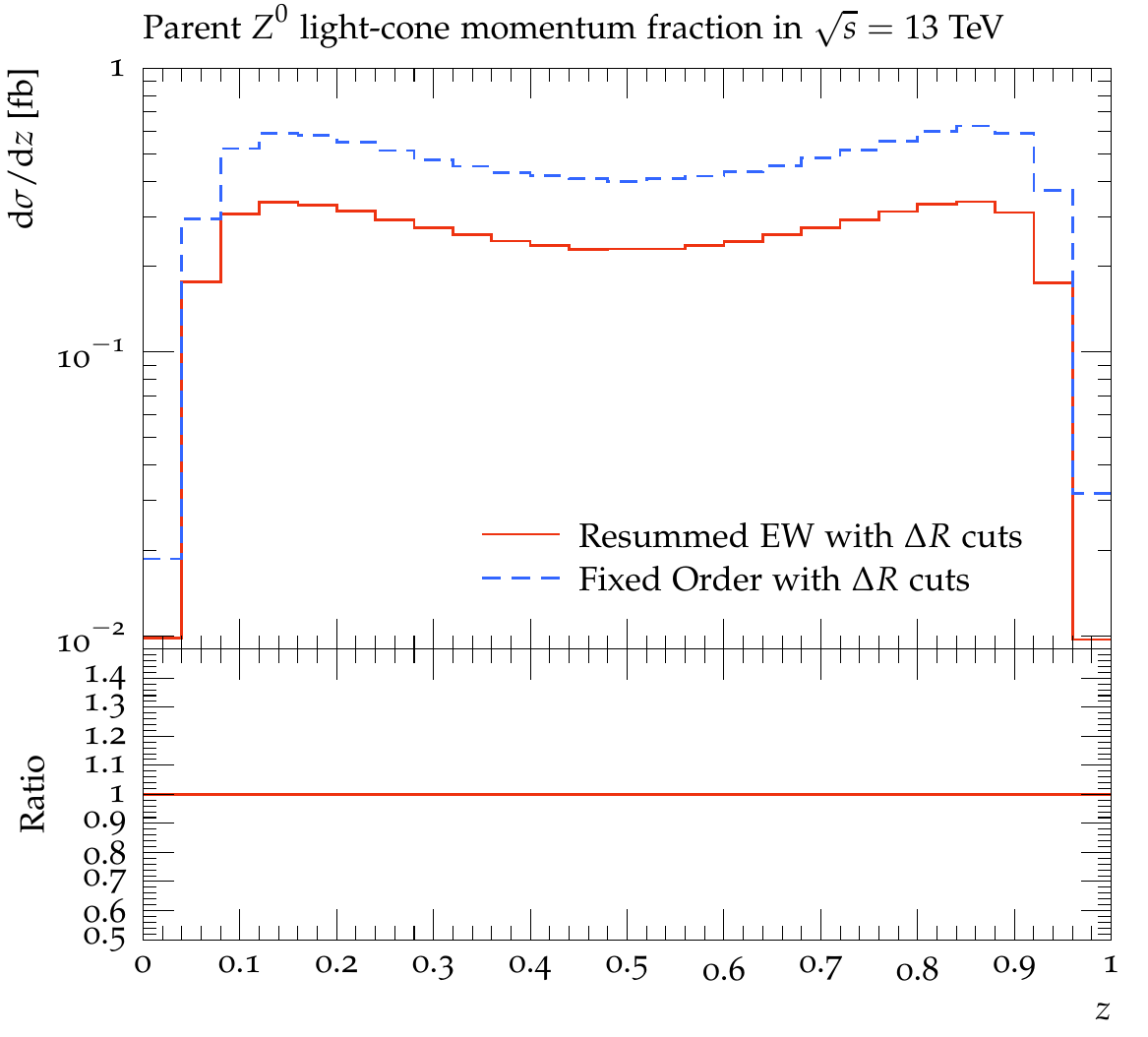}
  \caption{} 
  \label{}
\end{subfigure}
\caption{Performance test for $Z^0 \to W^{+} W^{-}$ EW branching in \textsf{Herwig 7} for $\sqrt{s}=13$ TeV. The panel (a) shows the differential rate of $Z^0$ emissions as functions of its transverse momentum while the panel (b) demonstrates the distribution of the light-cone momentum fraction of the parent bosons, $W^{\pm}$. The red histograms are obtained from showering a $q+\bar{q}\to W^{\pm}+Jet$ event with EW radiations, limited to a sngle radiation. The blue dashed histograms correspond to the equivalent fixed order events depicted at Figure \ref{FO-VVV-diags}. In both cases, to suppress the contributions that come from the channels described in diagrams (c), (d) and (e) of Figure~\ref{FO-VVV-diags} we have imposed the angular separation cuts (\ref{DRcuts}) and the invariant mass cut (\ref{InvMCut}). }
\label{zww-13TeV}
\end{figure}

\begin{figure}
\begin{subfigure}{.5\textwidth}
  \centering
  \includegraphics[width=1\linewidth]{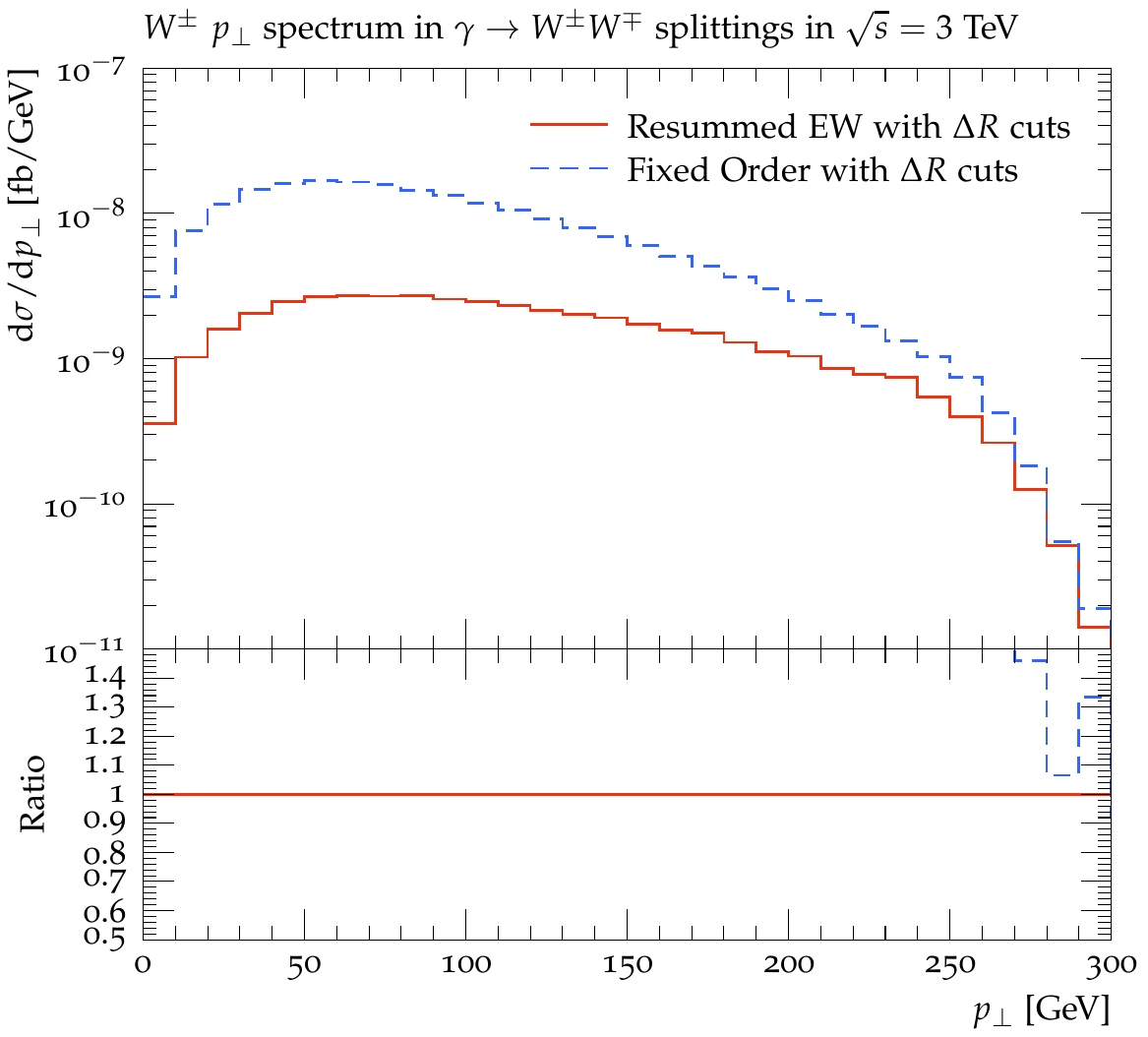}
  \caption{}
  \label{}
\end{subfigure}%
\begin{subfigure}{.5\textwidth}
  \centering
  \includegraphics[width=1\linewidth]{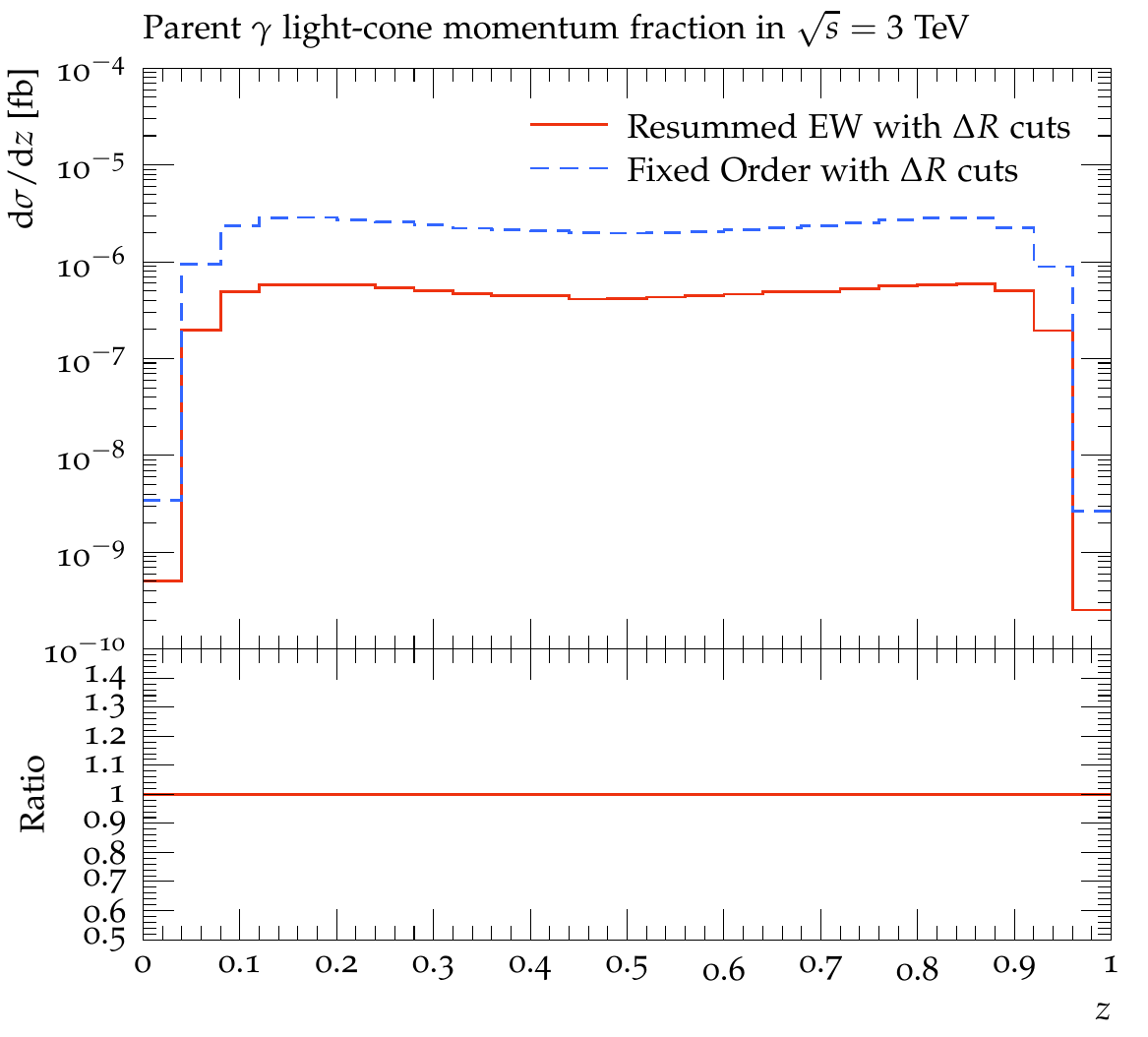}
  \caption{} 
  \label{}
\end{subfigure}
\caption{Performance test for $\gamma \to W^{+} W^{-}$ EW branching in \textsf{Herwig 7} for $\sqrt{s}=3$ TeV. A lower energy scale is used to suppress the $Z^0 \to W^{+} W^{-}$ events in the FO calculations. The notation of the plot is the same as in Figure~\ref{zww-13TeV}.}
\label{gww-3TeV}
\end{figure}

In the cases of the $Z^0 \to W^+ W^-$ and the $\gamma \to W^+ W^-$ branchings, we use a $q \bar{q} \to Z^0/\gamma + g$ underlying event in a similar fashion as in the previous cases and demonstrate the results in Figures~\ref{zww-13TeV} and \ref{gww-3TeV}. A particular issue on this analysis is that since these EW splittings produce a similar FS, one cannot separate these splittings in the corresponding FO calculation. To reduce this interference, we make use of an invariant mass cut,
\begin{eqnarray}
\hat{M} &=& \left[ \left(E_{W^+}+E_{W^-}+E_{\rm jet} \right) ^2 - \left| \bold{p}_{W^+}+\bold{p}_{W^-}+\bold{p}_{\rm jet} \right|^2 \right]^{1/2} > 2 \; {\rm TeV}. 
\label{InvMCut}
\end{eqnarray}
Nevertheless, the existence of the above-mentioned interference is clearly reflected in the distributions of the light-cone momentum fraction of the parent bosons for each splitting (see panels (b) of Figures~\ref{zww-13TeV} and \ref{gww-3TeV}), where there is a clear gap between the kinematics of the parent gauge bosons in the EW resummed and the FO results. Furthermore, since the contributions coming from the $Z^0 \to W^+ W^-$ vertex would be dominant in higher energy scales, the performance test of the $\gamma \to W^+ W^-$ EW branching has been done for $\sqrt{s}=3$ TeV. With all these considerations in place, we can observe that the implemented EW shower algorithm does a decent job of reproducing the FO results. 

\begin{figure}
\begin{subfigure}{.5\textwidth}
  \centering
  \includegraphics[width=1\linewidth]{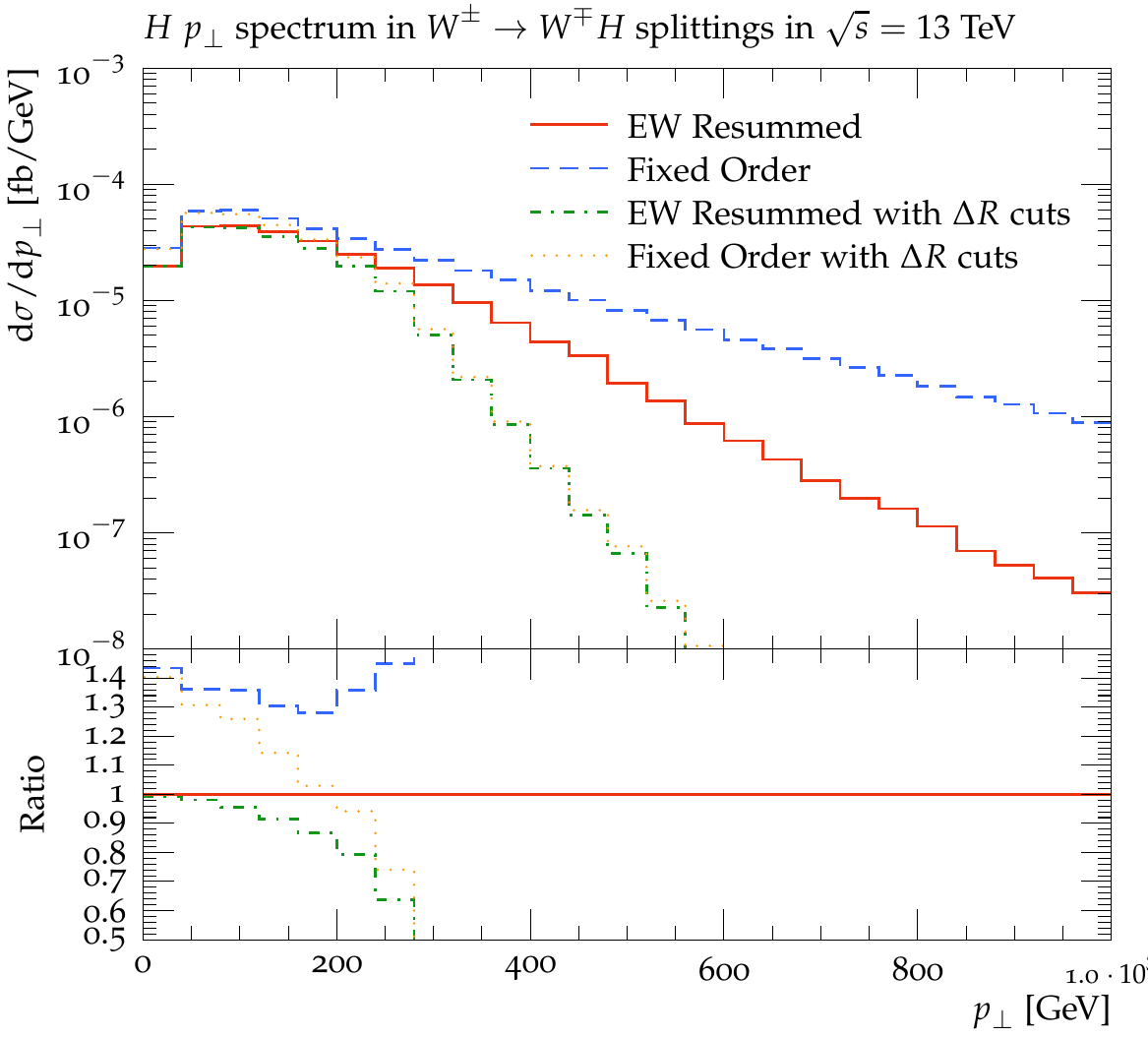}
  \caption{}
  \label{}
\end{subfigure}%
\begin{subfigure}{.5\textwidth}
  \centering
  \includegraphics[width=1\linewidth]{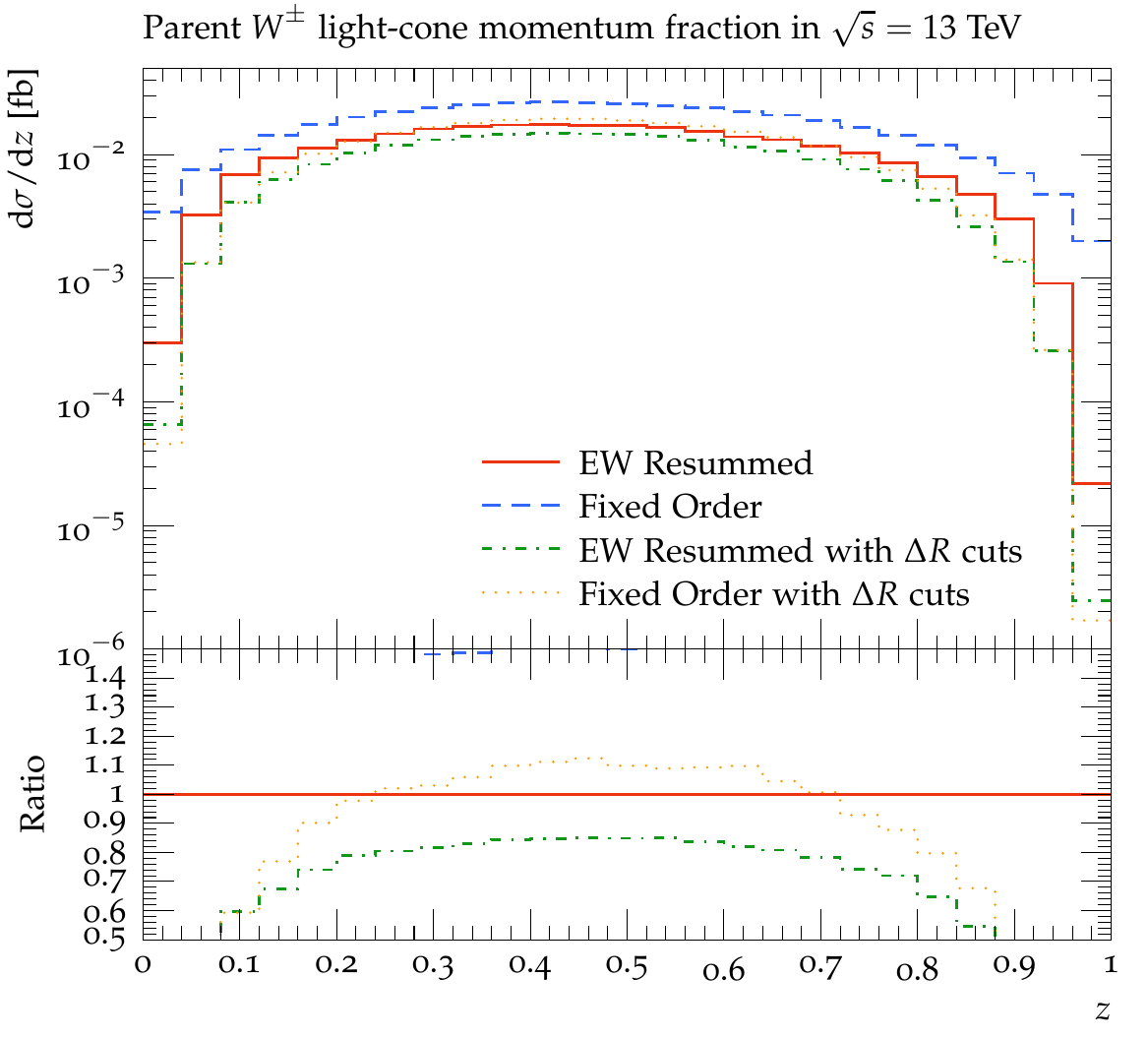}
  \caption{} 
  \label{}
\end{subfigure}
\caption{Performance test for $W^{\pm} \to W^{\pm} H$ EW branching in \textsf{Herwig 7} for $\sqrt{s}=13$ TeV. The notation of the plot is the same as in Figure~\ref{wwz-13TeV}.}
\label{wwh-13TeV}
\end{figure}

\begin{figure}
\begin{subfigure}{.5\textwidth}
  \centering
  \includegraphics[width=1\linewidth]{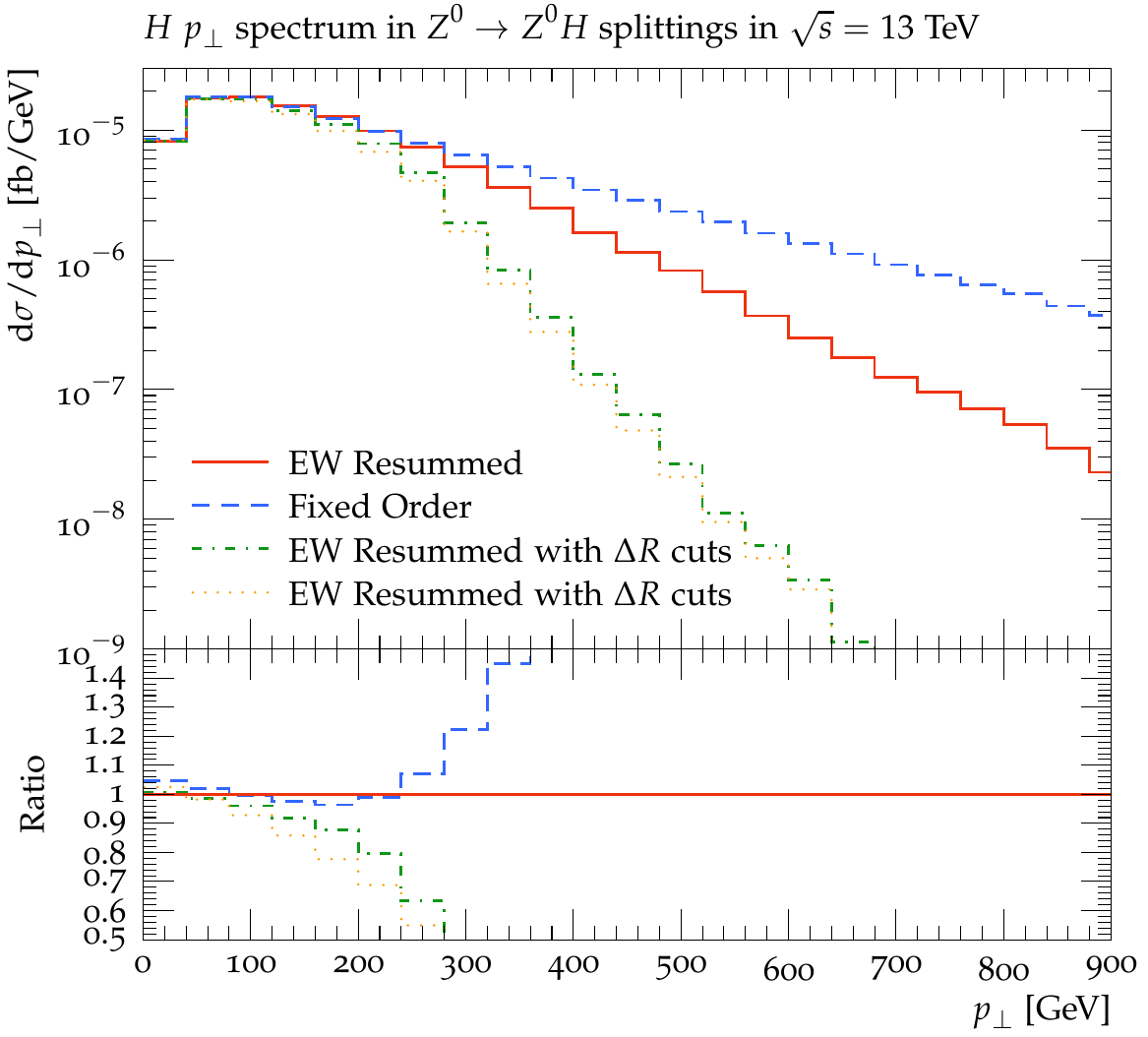}
  \caption{}
  \label{}
\end{subfigure}%
\begin{subfigure}{.5\textwidth}
  \centering
  \includegraphics[width=1\linewidth]{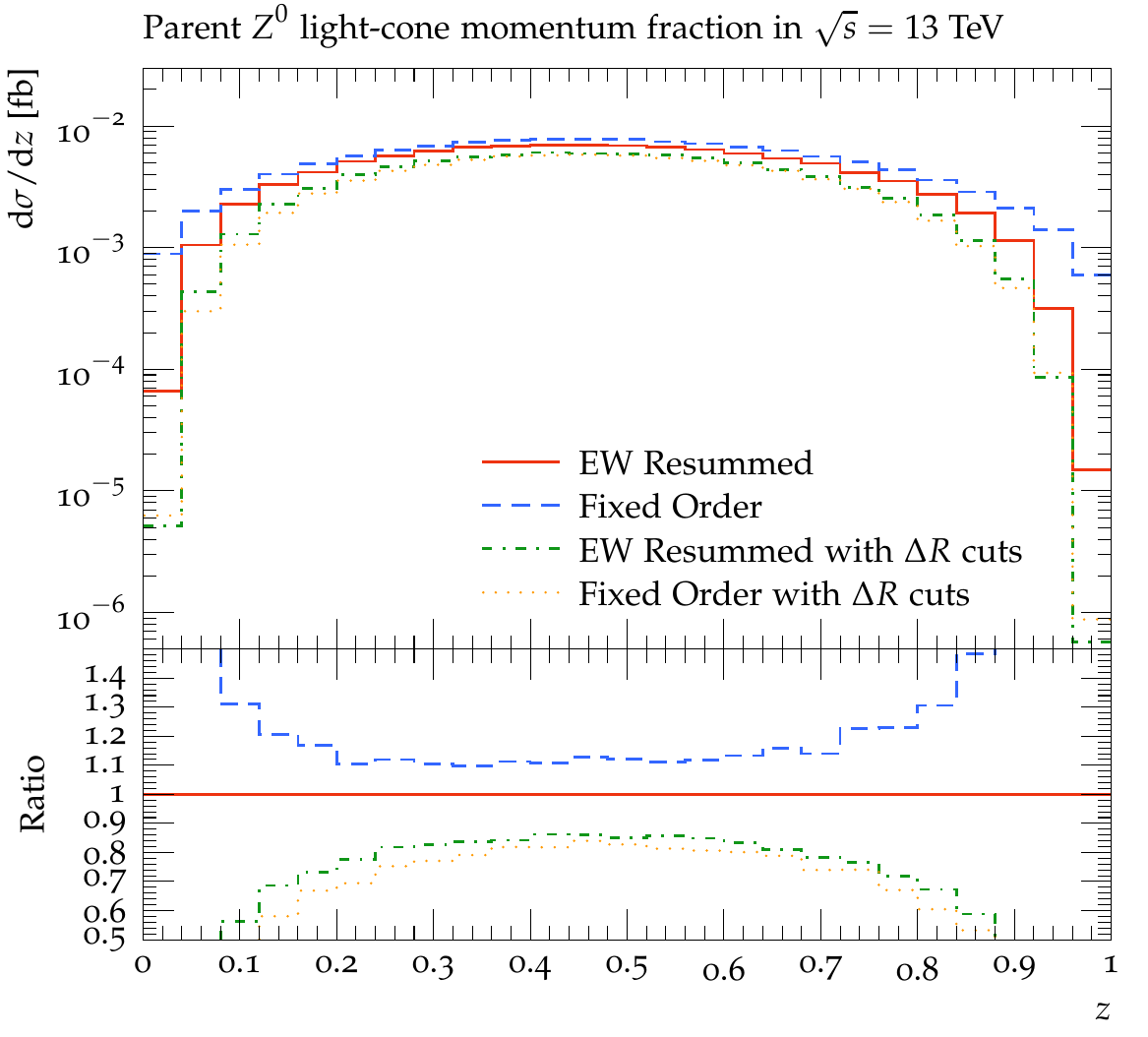}
  \caption{} 
  \label{}
\end{subfigure}
\caption{Performance test for $Z^0 \to Z^0 H$ EW branching in \textsf{Herwig 7} for $\sqrt{s}=13$ TeV. The notation of the plot is the same as in Figure~\ref{wwz-13TeV}.}
\label{zzh-13TeV}
\end{figure}

The last set of tests that we present here, are designed to check the performances of the $W^{\pm} \to W^{\mp} H$ and $Z^{0} \to Z^{0} H$ EW branchings, using a $q \bar{q} \to V + g$ underlying event. We also use the channels shown in Figure~\ref{FO-VVV-diags}, replacing one of the FS gauge bosons with a Higgs boson. One should note that in these cases, diagrams (c), (d) and (f) are strongly suppressed, since we are limiting the incoming quark flavors to \textit{up} and \textit{down}. The results are shown in Figures~\ref{wwh-13TeV} and \ref{zzh-13TeV}. 

\begin{figure}
\centering
\includegraphics[width=.6\textwidth]{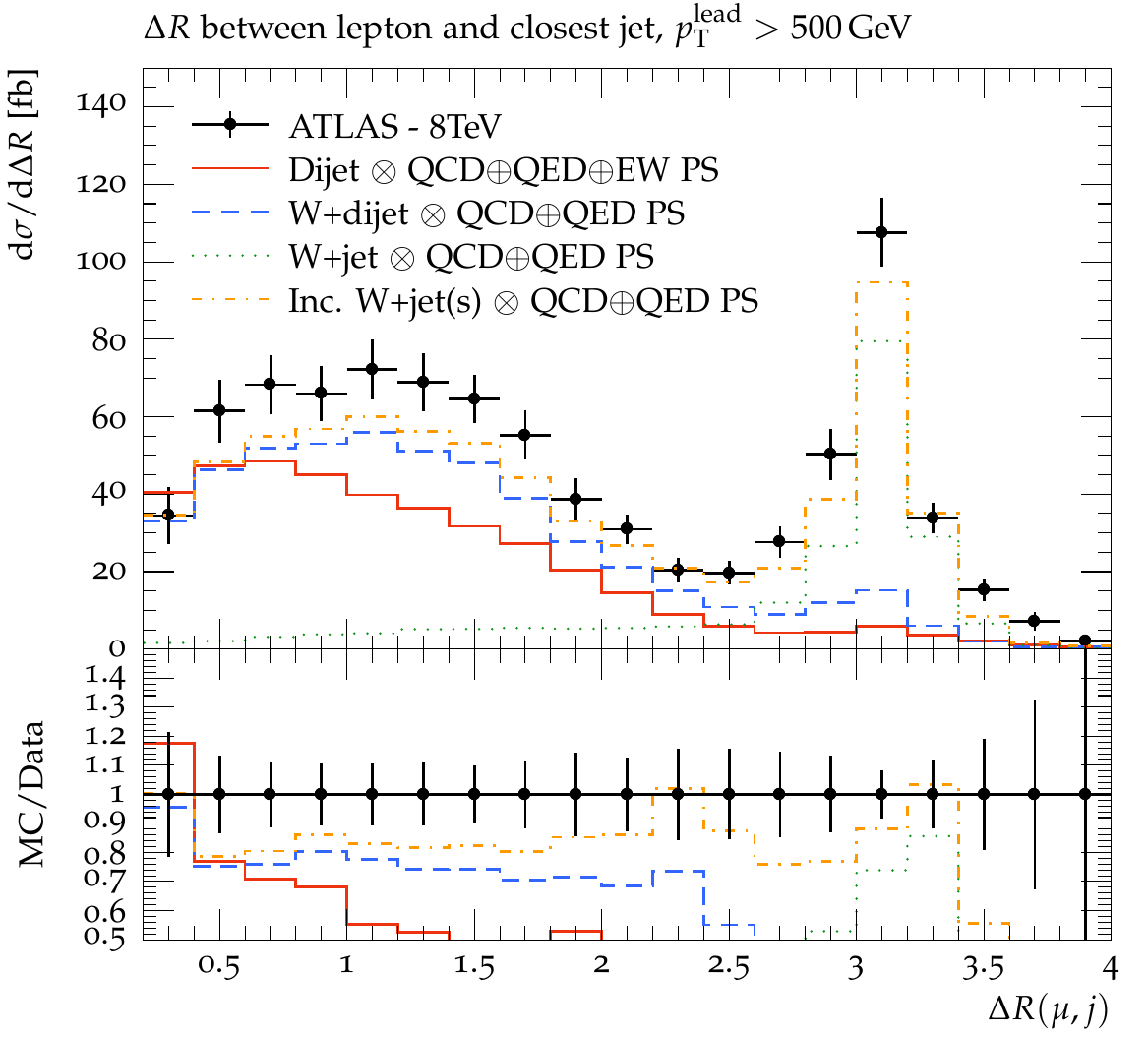}
\caption{The angular distribution of $W^{\pm}$ bosons accompanied with high transverse momentum jets at $\sqrt{s}=8$ TeV. The data is from ATLAS \cite{Aaboud:2016ylh}, showing the angular distribution of the muon and the closest jet with $p_t > 500$ GeV. The blue dashed and green dotted histograms are produced using $W^{\pm}+jet$ and $W^{\pm}+2jets$ MEs, respectively, while the orange dash-dotted histogram represents their sum. The corresponding MEs have been generated via \textsf{MadGraph} and showered in \textsf{Herwig} with its $QCD+QED$ scheme. On the other hand, the red solid histogram is calculated by showering a pure QCD dijet event with \textsf{Herwig}'s new $QCD+QED+EW$ scheme.}
\label{ATLAS-500}
\end{figure}

\begin{figure}
\centering
\includegraphics[width=.6\textwidth]{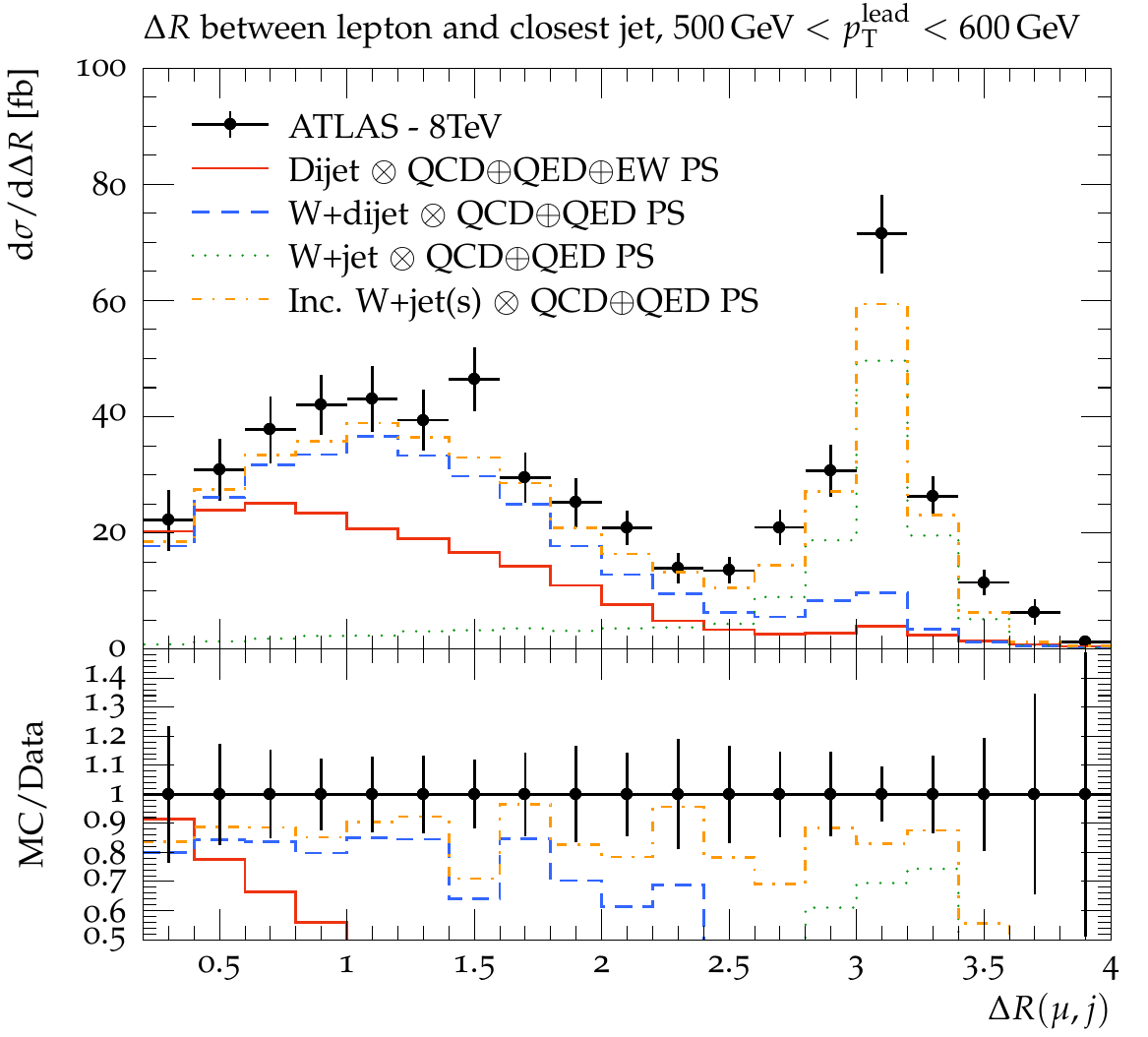}
\caption{The angular distribution of W bosons accompanied with high transverse momentum jets at $\sqrt{s}=8$ TeV. The data is from ATLAS \cite{Aaboud:2016ylh}, showing the angular distribution of the muon and the closest jet with $ 500 \; {\rm GeV} < p_t < 650 \; {\rm GeV}$. The notation of the plot is the same as in Figure \ref{ATLAS-500}.}
\label{ATLAS-500650}
\end{figure}

\begin{figure}
\centering
\includegraphics[width=.6\textwidth]{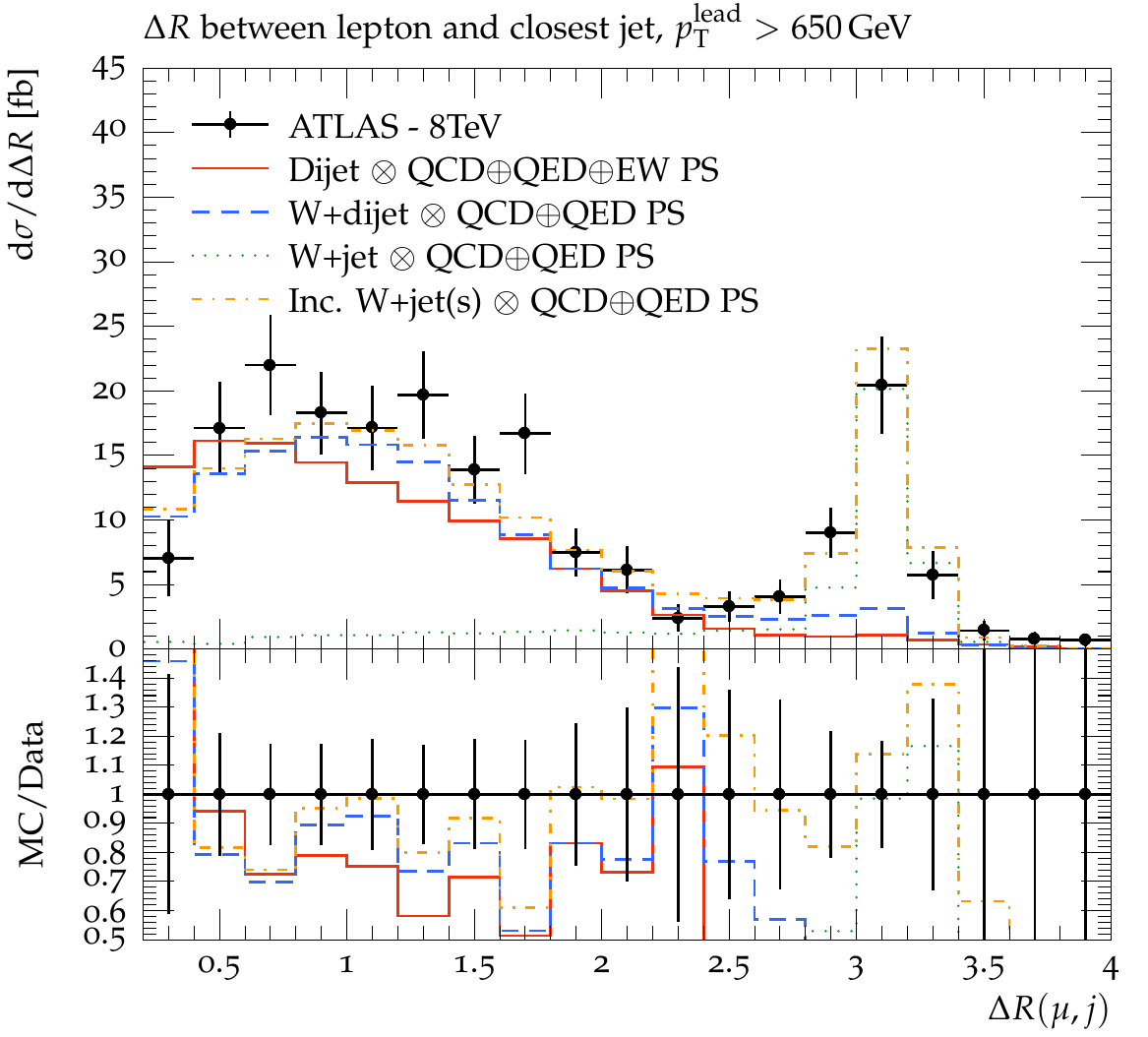}
\caption{The angular distribution of W bosons accompanied with high transverse momentum jets at $\sqrt{s}=8$ TeV. The data is from ATLAS \cite{Aaboud:2016ylh}, showing the angular distribution of the muon and the closest jet with $p_t > 650 \; {\rm GeV}$. The notation of the plot is the same as in Figure~\ref{ATLAS-500}.}
\label{ATLAS-650}
\end{figure}

Finally, after conforming that the implemented EW parton shower in \textsf{Herwig 7} can soundly describe the corresponding FO events, we can move on and use this EW shower in some physics tests. To this end, we calculate the angular distribution of $W^{\pm}$ bosons accompanied with high-transverse-momentum jets at $\sqrt{s}=8$ TeV. Such a measurement has been done by the ATLAS collaboration \cite{Aaboud:2016ylh}, reporting the angular separations of the observed muons and the closest jet with $p_t^{\rm leading \; jet} > 500$ GeV. Henceforth, we have generated $W^{\pm}+jet$ and $W^{\pm}+2jets$ MEs by \textsf{MadGraph5} and showered them with the \textit{QCD+QED} parton shower scheme in \textsf{Herwig 7}. These results are presented in Figures~\ref{ATLAS-500}, \ref{ATLAS-500650} and \ref{ATLAS-650}, respectively corresponding to $p_t^{\rm leading \; jet} > 500$ GeV, $500 \; {\rm GeV}> p_t^{\rm leading \; jet} > 560$ GeV and $p_t^{\rm leading \; jet} > 650$ GeV, where the $W^{\pm}+jet$ and $W^{\pm}+2jets$ calculations are plotted separately and their total values (labelled as \texttt{Inc.W+jet(s) $\otimes$ QCD$\oplus$QED PS}) are compared against the relevant ATLAS data. 

On the other hand, to get a sense of how effective the implemented EW parton shower is, we use \textsf{Herwig}'s internal ME, \texttt{MEQCD2to2}, to produce a pure QCD dijet event and set the minimum transverse momenta of these jets to be $500$ GeV. These events can be showered with \textsf{Herwig}'s new $QCD+QED+EW$ scheme. The results are plotted with red solid histograms in Figures~\ref{ATLAS-500}, \ref{ATLAS-500650} and \ref{ATLAS-650}, labelled as \texttt{Dijet $\otimes$ QCD$\oplus$QED$\oplus$EW PS}. Expectedly, since the corresponding events do not include explicit prompt $W^{\pm}$ emissions, they fall short of the \texttt{W$^{\pm}$+dijet $\otimes$ QCD$\oplus$QED PS} contributions. Nevertheless, the fact that our QCD dijet plus EW parton shower framework can closely describe the behavior of the $W^{\pm}+2jets$ calculations, shows the capability and soundness of the implemented EW parton shower in \textsf{Herwig 7}.

\section{Conclusion}
\label{sec:conc}

In the present work, we have outlined the necessary steps that are required for implementing an AO EW parton shower scheme in \textsf{Herwig 7} that includes IS EW radiations and an all-inclusive FS EW parton shower. We have systematically introduced all relevant \textit{quasi}-collinear EW splitting functions and derived their explicit helicity dependant forms. Afterward, these functions have been implemented in the \textsf{Herwig 7} AO shower algorithm, upgrading its existing \textit{QCD+QED} scheme to a new \textit{QCD+QED+EW} scheme. In the next step, we have run a comprehensive performance test, checking the implemented EW parton shower against the corresponding FO analysis by showering some relevantly suitable underlying events. This step has shown that our EW shower soundly describes successive EW radiations within the confinements of the collinear factorisation approximation. 

Furthermore, we have used our EW parton shower, as a part of \textsf{Herwig}'s \textit{QCD+QED+EW}  parton shower scheme, to simulate an LHC high-energy event, \textit{i.e.} the calculation of the angular distribution of $W^{\pm}$ bosons accompanied with high-transverse-momentum jets at $\sqrt{s}=8$ TeV. To do so, we have showered a purely QCD dijet production event with the \textit{QCD+QED+EW} shower and compared the results against inclusive $W^{\pm}+\rm jet(s)$ events and the existing experimental data from ATLAS. It has been shown that our simplistic framework, although being deprived of hard $W^{\pm}$ radiations, can predict the behaviour of the targeted events. 

The developed EW parton shower scheme would be available to the public, with the \textsf{Herwig~7.3} public release.

\section*{Acknowledgements}
\noindent
We thank our fellow Herwig authors for useful discussions and especially thank S.~Plätzer for reviewing this paper. This work has received funding from the European Union’s Horizon 2020 research and innovation program as part of the Marie Skłodowska-Curie Innovative Training Network MCnetITN3 
(grant agreement no.~722104). \textit{MRM} is also supported by the UK Science and Technology Facilities Council (grant numbers~ST/P001246/1).

\appendix

\section{EW shower in Herwig Interface}
\label{sec:HI}

After \textsf{Herwig-7.3} public release, the EW AO shower would be switched on in all input files by default. If not provided by default, or in a case where a change in the shower setting is needed, a shower switch option can be explicitly used by setting up the \texttt{ShowerHandler} interface as:

\mybox{\centering \texttt{set /Herwig/Shower/ShowerHandler:Interactions ALL}}

Here, the option \texttt{ALL} corresponds to the \textit{QCD+QED+EW} shower scheme. The other available options are \texttt{QEDQCD}, \texttt{QCD}, \texttt{QED} and \texttt{EWOnly}. One should note that, technically speaking, the $\gamma \to W^+ W^-$ EW branching is implemented a part of the QED parton shower and is accessible through either \texttt{ALL} or \texttt{QED} options. 

Besides the above options, all other accepts of the newly implemented parton shower is similar to the previous AO shower and can be found in \cite{Bahr:2008pv} or on the \textsf{Herwig} project webpage\footnote{https://herwig.hepforge.org/tutorials/showers/qtilde.html}.

\newpage

\begin{figure}
\begin{subfigure}{.5\textwidth}
  \centering
  \includegraphics[width=1\linewidth]{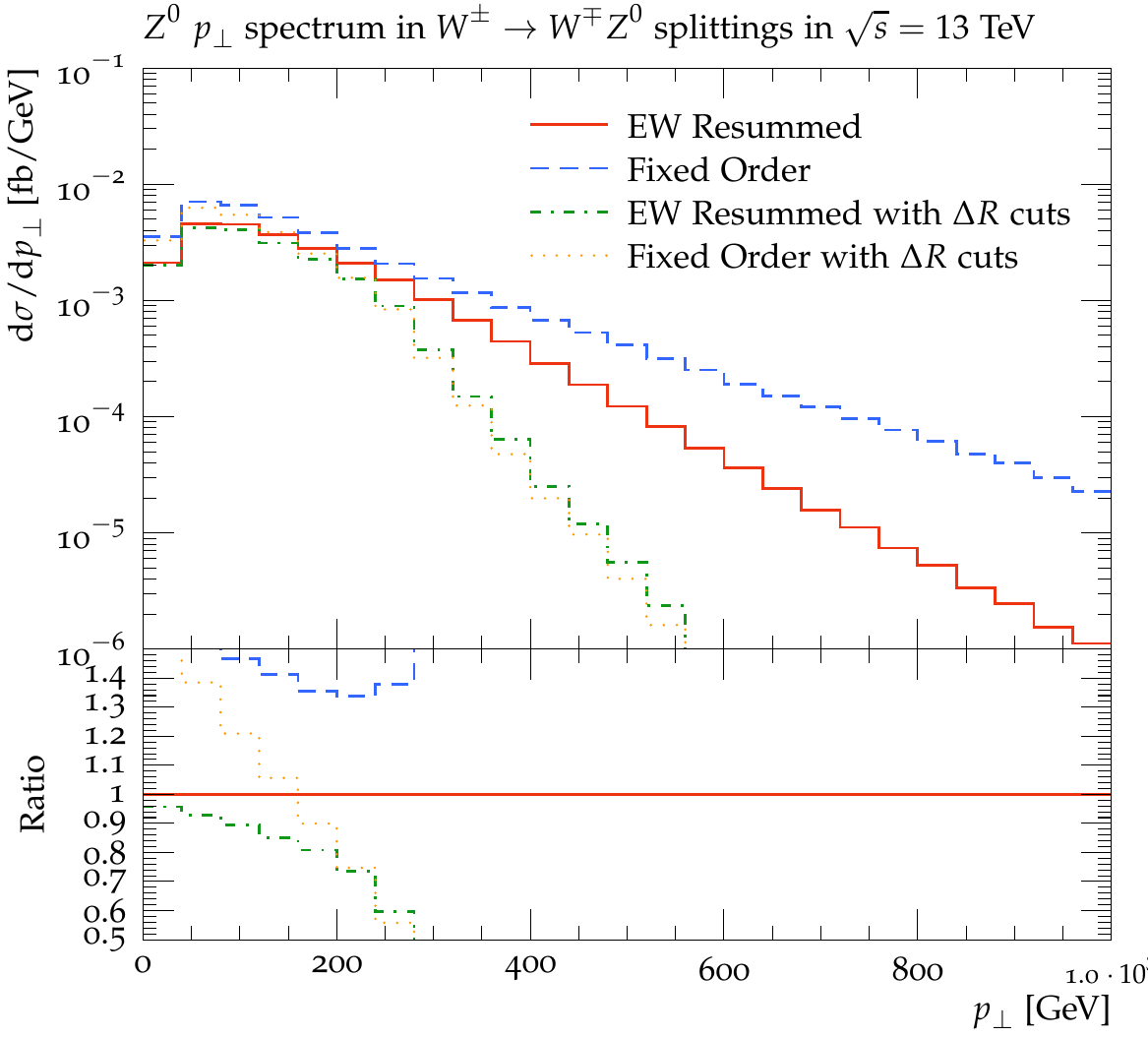}
  \caption{}
  \label{}
\end{subfigure}%
\begin{subfigure}{.5\textwidth}
  \centering
  \includegraphics[width=1\linewidth]{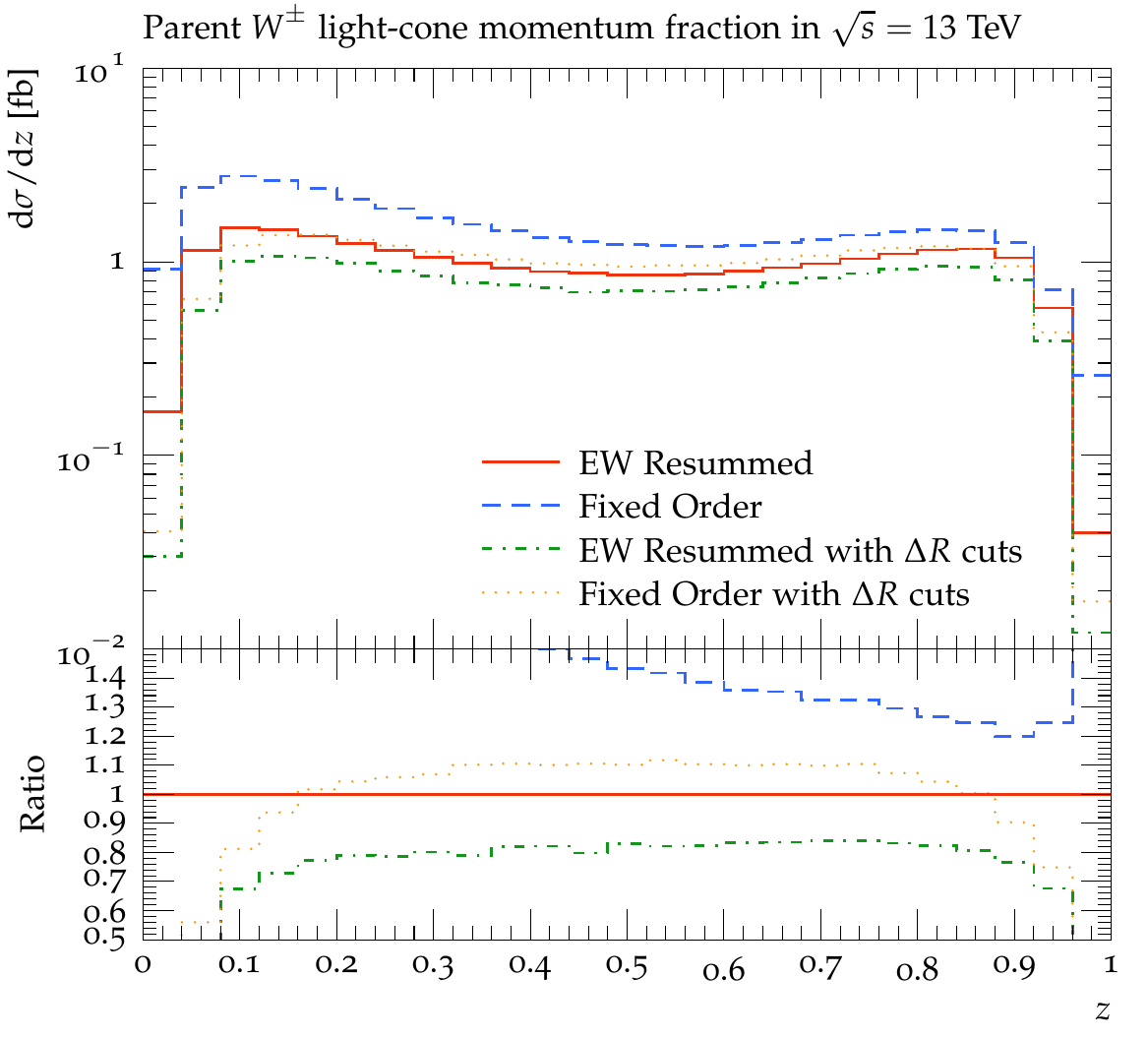}
  \caption{} 
  \label{wwz-13TeV}
\end{subfigure}
\caption{Performance test for $W^{\pm} \to W^{\pm} Z^0$ EW branching in \textsf{Herwig 7} for $\sqrt{s}=13$ TeV. The panel (a) shows the differential rate of $Z^0$ emissions as functions of its transverse momentum while the panel (b) demonstrates the distribution of the light-cone momentum fraction of the parent bosons, $W^{\pm}$. The red histograms are obtained from showering a $q+\bar{q}\to W^{\pm}+Jet$ event with EW radiations, limited to a sngle radiation. The blue dashed histograms correspond to the equivalent fixed order events depicted at Figure \ref{FO-VVV-diags}. In order to suppressing the contributions that come from the channels described in diagrams (c), (d) and (e) of Figure~\ref{FO-VVV-diags} we have imposed the angular separation cuts (\ref{DRcuts}) and plotted the results with green dash-dotted and orange dotted histograms for the EW single-step resmmation and FO calculations, respectively.}
\label{wwz-13TeV}
\end{figure}

\end{document}